\documentclass[3p,authoryear,12pt]{elsarticle}
\usepackage{amsthm}
\usepackage{amsmath}
\usepackage{amssymb}
\usepackage{xcolor}

\newcommand{\pl}{\partial}
\newcommand{\mc}{\mathcal}
\newcommand{\mf}{\mathbf}

\newcommand{\D}{\ensuremath{\mathcal{D}}}
\newcommand{\Id}{\ensuremath{\mathcal{I}}}

\newtheorem{thm}{Theorem}
\newdefinition{defn}{Definition}
\newdefinition{expl}{Example}
\newdefinition{step}{Step}
\newproof{pf}{Proof}

\begin{document}
\begin{frontmatter}

\title{Symbolic computation of conservation laws for nonlinear partial
differential equations in multiple space dimensions}

\thanks{The research was supported in part by the National Science Foundation
under Grant No.\ CCF-0830783. }

\author{Douglas Poole and Willy Hereman}
\address{Department of Mathematical and Computer Sciences,
Colorado School of Mines, Golden, CO 80401, USA}
\ead{whereman@mines.edu and lpoole@mines.edu}
\ead[url]{http://inside.mines.edu/~whereman}
\begin{abstract}
A method for symbolically computing conservation laws of nonlinear partial
differential equations (PDEs) in multiple space dimensions is presented in
the language of variational calculus and linear algebra.
The steps of the method are illustrated using the Zakharov-Kuznetsov and
Kadomtsev-Petviashvili equations as examples.

The method is algorithmic and has been implemented in {\it Mathematica}.
The software package, {\sc ConservationLawsMD.m}, can be used to symbolically
compute and test conservation laws for polynomial PDEs that can 
be written as nonlinear evolution equations.

The code {\sc ConservationLawsMD.m} has been applied to multi-dimensional 
versions of the Sawada-Kotera, Camassa-Holm, Gardner, and 
Khokhlov-Zabolotskaya equations.
\end{abstract}


\end{frontmatter}
%
\section{Introduction}
\label{secIntroduction}
Many nonlinear partial differential equations (PDEs) in the applied sciences 
and engineering are {\it continuity equations} which express conservation 
of mass, momentum, energy, or electric charge.
Such equations occur in, e.g., fluid mechanics, particle and quantum physics, 
plasma physics, elasticity, gas dynamics, electromagnetism, 
magneto-hydro-dynamics, nonlinear optics, etc.
Certain nonlinear PDEs admit infinitely many conservation laws. 
Although most lack a physical interpretation, these conservation laws play 
an important role in establishing the {\it complete integrability} of the PDE.
Completely integrable PDEs are nonlinear PDEs that can be linearized by some
transformation (e.g., the Cole-Hopf transformation linearizes the Burgers 
equation) or explicitly solved with the Inverse Scattering Transform (IST). 
See, e.g., \citet{AblowitzClarkson91}.

The search for conservation laws of the Korteweg-de Vries (KdV) equation
began around 1964 and the knowledge of conservation laws was paramount for 
the development of soliton theory. 
As \citet{Newell83} narrates, the study of conservation laws led to the 
discovery of the Miura transformation (which connects solutions of the KdV 
and modified KdV (mKdV) equations) and the Lax pair \citep{Lax68}, 
i.e., a system of linear equations which are only compatible if the original 
nonlinear PDE holds.
In turn, the Lax pair is the starting point for the IST 
\citep{AblowitzClarkson91,AblowitzSegur81} which has been used to construct 
soliton solutions, i.e., stable solutions that interact elastically upon 
collision.

Conversely, the existence of many (independent) conserved densities is a 
{\it predictor} for complete integrability.
The knowledge of conservation laws also aids the study of qualitative 
properties of PDEs, in particular, bi-Hamiltonian structures and recursion 
operators \citep{BaldwinHereman10}.
Furthermore, if constitutive properties have been added to close ``a model,''
one should verify that conserved quantities have remained intact. 
Another application involves numerical solvers for PDEs \citep{Sanzserna82}, 
where one checks if the first few (discretized) conserved densities are 
preserved after each time step.

There are several methods for computing conservation laws as discussed 
by e.g., 
\vskip 1pt
\noindent
\citet{Blumanetal10}, \citet{HeremanAll05}, 
\citet{Naz08}, \citet{Nazetal08}, and \citet{Rosenhaus02}.
One could apply Noether's theorem, which states that a (variational) 
symmetry of the PDE corresponds to a conservation law.
Using Noether's method, the {\sc DifferentialGeometry} package in {\it Maple} 
contains tools for conservation laws developed by 
\citet{Andersonsoftware04} and \citet{AndersonChebTerMaple09}. 
Circumventing Noether's theorem, \citet{Wolf02} has developed four programs 
in {\sc REDUCE} which solve an over-determined system of differential
equations to get conservation laws. 
Based on the integrating factor method, \citet{Cheviakov07,Cheviakov10} 
has written a {\it Maple} program that computes a set of integrating 
factors (multipliers) on the PDE. 
To find conservation laws, here again, one has to solve a system of 
differential equations.
The {\it Maple} package PDEtools by \citet{ChebTerrabvonBulowMaple04} has 
the commands {\sc ConservedCurrents} and {\sc ConservedCurrentTest} for 
computing and testing conservation laws using the integrating factor method.
Last, conservation laws can be obtained from the Lax operators, as shown by, 
e.g., \citet{ZakharovShabat72} and \citet{DrinfeldSokolov85}.

By contrast, the method discussed in this paper uses tools from calculus, 
the calculus of variations, linear algebra, and differential geometry.
Briefly, our method works as follows.
A candidate (local) density is assumed to be a linear combination with 
undetermined coefficients of monomials that are invariant under the 
scaling symmetry of the PDE.
Next, the time derivative of the candidate density is computed and evaluated 
on the PDE. 
Subsequently, the variational derivative is applied to get a linear system 
for the undetermined coefficients.
The solution of that system is substituted into the candidate density.
Once the density is known, the flux is obtained by applying a homotopy 
operator to invert a divergence.
Our method can be implemented in any major computer algebra system (CAS).
The package {\sc ConservationLawsMD.m} by \citet{PooleHereman09software2} is
a {\it Mathematica} implementation based on work by \citet{HeremanAll05},
with new features added by \citet{Poole09}.

This paper is organized as follows.
To set the stage, Section~\ref{secExamples} shows conservation laws for the 
Zakharov-Kuznetsov (ZK) and Kadomtsev-Petviashvili (KP) equations.
Section~\ref{secNotationTools} covers the tools that will be used 
in the algorithm.
In Section~\ref{secAlgorithm}, the algorithm is presented and illustrated 
for the ZK and KP equations.
Section~\ref{secApplications} discusses conservation laws of PDEs in 
multiple space dimensions, including the Khokhlov-Zabolotskaya (KZ) 
equation and multi-dimensional versions of the Sawada-Kotera (SK),  
Camassa-Holm (CH) and Gardner equations.
Conservation laws for the multi-dimensional SK, CH, and Gardner equations 
were not found in a literature survey and are presented here for the first 
time.
A general conservation law for the KP equation is given in 
Section~\ref{secKPequation}.
Using the (2+1)-dimensional Gardner equation as an example, 
Section~\ref{secUsingProgram} shows how to use {\sc ConservationLawsMD.m}.
Finally, some conclusions are drawn in Section~\ref{secConclusion}.
%
\section{Examples of Conservation Laws}
\label{secExamples}
This paper deals with systems of polynomial PDEs of order $M,$
\begin{equation}
\label{definitionPDEonevar}
\mf{\Delta}(\mf{u}^{(M)}(\mf{x})) = \mf{0},
\end{equation}
in $n$ dimensions where $\mf{x} = (x^1,x^2,\dots,x^n)$ is the independent 
variable. $\mf{u}^{(M)}(\mf{x})$ denotes the dependent variable 
$\mf{u} = (u^1, \dots, u^j, \dots, u^N)$ and its partial derivatives 
(up to order $M$) with respect to $\mf{x}.$ 
We do not cover systems of PDEs with variable coefficients.

A conservation law for (\ref{definitionPDEonevar}) is a scalar PDE in the form
\begin{equation}
\label{divconlaw}
\mathrm{Div}\,\mf{P} = 0 
\mbox{\hspace{3mm} {\rm on} $\mf{\Delta} = \mf{0}$},
\end{equation}
where $\mf{P} = \mf{P}(\mf{x},\mf{u}^{(P)}(\mf{x}))$ of some order $P.$
The definition follows \citet{Olver93} and \citet{Blumanetal10}, and is
commonly used in literature on symmetries of PDEs.
In physics, $P$ is called a conserved current. 
More precisely, a conservation law can be viewed as an equivalence class of 
conserved currents \citep{Vinogradov89}.
Our algorithm computes one member from each equivalence class; usually a 
representative that is of lowest complexity and free of curl terms.

Since we work on PDEs from the physical sciences, the algorithm and code 
are restricted to 1\,D, 2\,D, and 3\,D in space, but can be extended to 
$n$ dimensions.
Indeed, many of our applications model dynamical problems,
where $\mf{x} = (x,y,t)$ for PDEs in 2\,D or $\mf{x} = (x,y,z,t)$ for PDEs 
in 3\,D in space.
In either case, the additional variable, $t,$ denotes time. 

Throughout the paper, we will use an alternative definition for 
(\ref{divconlaw}), 
\begin{equation}
\label{genconlaw}
\D_t \rho + \mathrm{Div} \,\mathbf{J} = 0 
\mbox{\hspace{3mm} {\rm on} $\mf{\Delta} = \mf{0}$},
\end{equation}
where $\rho = \rho(\mf{x},\mf{u}^{(K)}(\mf{x}))$ is the conserved
density of some order $K,$ and
$\mathbf{J} = \mathbf{J} \left(\mf{x},\mf{u}^{(L)}(\mf{x}) \right)$
is the associated flux of some order $L$
\citep{MiuraGardnerKruskal68, AblowitzClarkson91}.
Comparing (\ref{divconlaw}) with (\ref{genconlaw}), it should be clear that 
$\mathbf{P} = (\rho,\mathbf{J})$ with $P = {\rm max}\{ K,L \}.$

For simplicity, in the examples we will denote the dependent variables
$u^1, u^2, u^3,$ etc., by $u, v, w,$ etc.
Partial derivatives are denoted by subscripts, e.g.,
$\frac{\pl^{k_1 + k_2 + k_3} u} {\pl x^{k_1}\pl y^{k_2} \pl z^{k_3}}$ 
is written as $u_{k_1 x \, k_2 y \, k_3 z},$ where the $k_i$ are non-negative
integers.
In (\ref{genconlaw}), $\mathrm{Div}\,\mf{J}$ is the {\it total} divergence 
operator, where 
$\mathrm{Div}\,\mf{J} = \D_x J^x + \D_y J^y$ if $\mf{J} = (J^x, J^y)$ 
and $\mathrm{Div}\,\mathbf{J} = \D_x J^x + \D_y J^y + \D_z J^z$ 
if $\mf{J} = (J^x, J^y, J^z).$
Logically, $\D_t,$ $\D_x,$ $\D_y,$ and $\D_z$ are total derivative operators.
For example, the total derivative operator $\D_x$ (in 1\,D) acting on
$f = f(x,t,\mf{u}^{(M)}(x,t))$ of order $M$ is defined as
\begin{equation}
\label{totalderivative1D}
\D_x f =
\frac{\pl f}{\pl x} + \sum_{j=1}^N \sum_{k=0}^{M_1^j} u_{(k+1) x}^j
\frac{\pl f}{\pl u_{k x}^j},
\end{equation}
where $M_1^j$ is the order of $f$ in component $u^j$ and
$M = {\rm max} \{ M_1^1, \ldots, M_1^N \}.$
The partial derivative $\frac{\pl }{\pl x}$ acts on any $x$ that appears 
explicitly in $f,$ but not on $u^j$ or any partial derivatives of $u^j.$
Total derivative operators in multiple dimensions are defined analogously
(see Section~\ref{secNotationTools}).

The algorithm described in Section~\ref{secAlgorithm} allows one to compute
local conservation laws for systems of nonlinear PDEs that can be written as 
evolution equations.  
For example, if $\mf{x} = (x,y,z,t),$ an evolution equation in variable $t$ 
has the form
\begin{equation}
\label{evolutiondefinition}
\mathbf{u}_t = \mathbf{G}(u^1,u_x^1,u_y^1,u_z^1,u_{2x}^1,u_{2y}^1,u_{2z}^1,
u_{xy}^1,\dots,u^N_{M_1^N x\,M_2^N y\,M_3^N z} ),
\end{equation}
where $\mf{G}$ is assumed to be smooth and $M_1^j, M_2^j,$ and $M_3^j$ are
the orders of component $u^j$ with respect to $x, y,$ and $z,$ respectively,
and $M$ is the maximum total order of all terms in the differential function.
Few multi-dimensional systems of PDEs are of the form 
(\ref{evolutiondefinition}).
However, it is often possible to obtain a systems of evolution equations by 
recasting a single higher-order equation into a system of first-order 
equations, sometimes in conjunction with a simple transformation. 
If necessary, our program internally interchanges independent variables to 
obtain (\ref{evolutiondefinition}), where time is the evolution variable.
However, that swap of variables is not used in this paper.
For a clearer description of the algorithm, we allow systems of evolution 
equations where {\it any} component of $\mf{x}$ can play the role of 
evolution variable.

We now introduce two well-documented PDEs together with some of their 
conservation laws.
These PDEs will be used in Section~\ref{secAlgorithm} to illustrate the 
steps of the algorithm.
\begin{expl}
The Zakharov-Kuznetsov (ZK) equation is an evolution equation that models 
three-dimensional ion-sound solitons in a low pressure uniform magnetized 
plasma 
\vskip 1pt
\noindent
\citep{ZakharovKuznetsov74}.
After re-scaling, it takes the form
\begin{equation}
\label{generalZKequation}
u_t + \alpha u u_x + \beta ( \Delta u )_x = 0,
\end{equation}
where $u(\mf{x}) = u(x,y,z,t), \alpha$ and $\beta$ are real parameters, 
and $\Delta = \frac{\partial^2}{\partial x^2}
+ \frac{\partial^2}{\partial y^2} + \frac{\partial^2}{\partial z^2}$ is the 
Laplacian in 3\,D.
The conservation laws for the (2+1)-dimensional ZK equation,
\begin{equation}
\label{ZKequation2D}
u_t + \alpha u u_x + \beta ( u_{2x} + u_{2y} )_x = 0,
\end{equation}
where $u(\mf{x}) = u(x,y,t),$ were studied by, e.g.,
\citet{ZakharovKuznetsov74}, \citet{Infeld85}, and \citet{Shivamoggi93}.
After correcting some of the results reported in \citet{Shivamoggi93}, 
the polynomial conservation laws of (\ref{ZKequation2D}) are
\begin{equation}
\label{ZK2Dconlaw1} 
\D_t ( u ) + \D_x ( \tfrac{1}{2} \alpha u^2 + \beta u_{2x} )
+ \D_y ( \beta u_{xy} ) = 0,
\end{equation}
which corresponds to the ZK equation itself, and
\newpage
\noindent
\begin{eqnarray}
\label{ZK2Dconlaw2} 
\!\!\!\!\!\!\!\!\!\!\!\!\!\!\!\!
&& \D_t \Big( u^2 \Big) 
\!+\! \D_x \Big( \tfrac{2}{3} \alpha u^3 - \beta (u_x^2 - u_y^2) 
+ 2 \beta u (u_{2x} + u_{2y} )\Big)
\!+\! \D_y \Big( \!\!-2 \beta u_x u_y \Big) = 0,
\\*[-5mm] 
\!\!\!\!\!\!\!\!\!\!\!\!\!\!\!\! \nonumber \\ 
\label{ZK2Dconlaw3} 
\!\!\!\!\!\!\!\!\!\!\!\!\!\!\!\!
&& \D_t \Big( u^3 - 3\tfrac{\beta}{\alpha} (u_x^2 + u_y^2) \Big)
\!+\! \D_x \Big( 3 u^2 (\tfrac{1}{4} \alpha u^2 + \beta u_{2x} )
- 6 \beta u (u_x^2 + u_y^2) + 3\tfrac{\beta^2}{\alpha} (u_{2x}^2 - u_{2y}^2)
\nonumber \\ 
\!\!\!\!\!\!\!\!\!\!\!\!\!\!\!\! 
&& \hspace*{5mm} {}
- 6\tfrac{\beta^2}{\alpha} (u_x (u_{3x} + u_{x 2y})
+ u_y (u_{2x y} + u_{3y})) \Big)
\!+\! \D_y \Big( 3 \beta u^2 u_{xy} 
+ 6\tfrac{\beta^2}{\alpha} u_{xy} (u_{2x} + u_{2y}) \Big) = 0,
\\*[-5mm] 
\!\!\!\!\!\!\!\!\!\!\!\!\!\!\!\! \nonumber \\ 
\label{ZK2Dconlaw4} 
\!\!\!\!\!\!\!\!\!\!\!\!\!\!\!\!
&& \D_t \Big( t u^2 \!-\! \tfrac{2}{\alpha} x u \Big)
\!+\! \D_x \Big( t(\tfrac{2}{3} \alpha u^3 \!-\! \beta (u_x^2 \!-\! u_y^2)
\!+\! 2 \beta u (u_{2x} \!+\! u_{2y})) 
\!-\! \tfrac{2}{\alpha} x (\tfrac{1}{2} \alpha u^2 \!+\! \beta u_{2x}) 
\!+\! 2 \tfrac{\beta}{\alpha} u_x  \Big) 
\nonumber \\ 
\!\!\!\!\!\!\!\!\!\!\!\!\!\!\!\! 
&& \hspace*{5mm} {} \!+\! \D_y \Big( \!\!-2 \beta (t u_x u_y
\!+\! \tfrac{1}{\alpha} x u_{xy}) \Big) = 0.
\end{eqnarray}
\end{expl}
Note that the fourth conservation law (\ref{ZK2Dconlaw4}) explicitly 
depends on $t$ and $x.$
\begin{expl}
The well-known (2+1)-dimensional Kadomtsev-Petviashvili (KP) equation, 
\begin{equation}
\label{KPequation}
(u_t + \alpha u u_x + u_{3x})_x + \sigma^2 u_{2y} = 0,
\end{equation}
for $u(x,y,t),$ describes shallow water waves with wavelengths much greater 
than their amplitude moving in the $x$-direction and subject to weak 
variations in the $y$-direction 
\vskip 1pt
\noindent
\citep{Kadomtsev70}.
The parameter $\alpha$ occurs after a re-scaling of the physical coefficients
and $\sigma^2 = \pm 1.$ 
Obviously, the KP equation is not an evolution equation. 
However, it can be written as an evolution system in space variable $y,$ 
\begin{equation}
\label{KPevolutionform}
u_y = v, \quad
v_y = - \sigma^2 (u_{tx} + \alpha u_x^2 + \alpha u u_{2x} + u_{4x}).
\end{equation}
Note that $\frac{1}{\sigma^2} = \sigma^2,$ and thus $\sigma^4 = 1.$
System (\ref{KPevolutionform}) instead of (\ref{KPequation}) will be used in 
Section~\ref{secAlgorithm}. 
{\sc ConservationLawsMD.m} has an algorithm that will identify an evolution 
variable and transform the given PDE into a system of evolution equations.

Equation (\ref{KPequation}) expresses conservation of momentum:
\begin{equation}
\label{KPeqnasconlaw}
\D_t ( u_x ) + \D_x ( \alpha uu_x + u_{3x} ) 
+ \D_y ( \sigma^2 u_y ) = 0.
\end{equation}
Other well-documented conservation laws \citep{Wolf02} are
\begin{eqnarray}
\label{KPgenconlaw1} 
&& \D_t \Big( f u \Big) 
\!+\! \D_x \Big( f (\tfrac{1}{2} \alpha u^2 + u_{2x}) + 
(\tfrac{1}{2} \sigma^2 f^{\prime} y^2 - f x) (u_t + \alpha uu_x + u_{3x}) \Big)
\nonumber \\ 
&& \hspace*{5mm} {} 
\!+\! \D_y \Big( (\tfrac{1}{2} f^{\prime} y^2 - \sigma^2 f x) u_y
- f^{\prime} y u \Big) = 0, \\ 
\label{KPgenconlaw2} 
&& \D_t \Big( f y u \Big) 
\!+\! \D_x \Big( f y (\tfrac{1}{2} \alpha  u^2 + u_{2x}) 
+ y (\tfrac{1}{6} \sigma^2 f^{\prime} y^2 - f x)(u_t + \alpha uu_x + u_{3x}) 
\Big)
\nonumber \\ 
&& \hspace*{5mm} {}
\!+\! \D_y \Big( y (\tfrac{1}{6} f^{\prime} y^2 - \sigma^2 f x) u_y
- (\tfrac{1}{2} f^{\prime} y^2 - \sigma^2 f x) u \Big) = 0,
\end{eqnarray}
where $f = f(t)$ is an arbitrary function.
Thus, there is an infinite family of conservation laws, each of the form 
(\ref{KPgenconlaw1}) or (\ref{KPgenconlaw2}).
In Section~\ref{secAlgorithm} we will show how 
(\ref{ZK2Dconlaw1})-(\ref{ZK2Dconlaw4}) are computed straightforwardly with 
our algorithm. 
We will also compute several conservation laws for the KP equation. 
Our current code does not (algorithmically) compute (\ref{KPgenconlaw1}) and 
(\ref{KPgenconlaw2}).
Instead, conservation laws obtained with the code allow the user to 
conjecture and test the form of (\ref{KPgenconlaw1}) and (\ref{KPgenconlaw2}).
In Section~\ref{secKPequation}, we give computational details and show how 
(\ref{KPgenconlaw1}) and (\ref{KPgenconlaw2}) can be verified.
\end{expl}
%
\section{Tools from the Calculus of Variations and Differential Geometry}
\label{secNotationTools}
Three operators from the calculus of variations and differential geometry 
play a major role in the conservation law algorithm.
Namely, the total derivative operator, and the Euler and homotopy operators.
All three operators (which act on the jet space) can be defined 
algorithmically which allows for straightforward and efficient computations.

The algorithm in Section~\ref{secAlgorithm} requires that operations 
applied to {\it differential functions} $f(\mf{x},\mf{u}^{(M)}(\mf{x})),$
take place in the jet space, where one component of $\mf{x}$ is a parameter.

Although in later sections, one of the space variables will serve as the 
parameter, in this section we arbitrarily choose $t$ as the parameter 
(matching (\ref{evolutiondefinition})).
Thus, in all definitions and theorems in this section, 
1\,D means that there is only one space variable, yet $\mf{x} = (x,t).$ 
Likewise, in 2\,D and 3\,D cases, $\mf{x} = (x,y,t)$ and 
$\mf{x} = (x,y,z,t),$ respectively.

Using (\ref{evolutiondefinition}), we assume that all partial derivatives 
of $\mf{u}$ with respect to $t$ are eliminated from $f.$ 
Thus, $f(\mf{x}, \mf{u}^{(M)}(\mf{x}))$ with
\begin{equation}
\label{jetspacecoord}
\mf{u}^{(M)}(\mf{x}) =
( u^1,u_x^1,u_y^1,u_z^1,u_{2x}^1,u_{2y}^1,u_{2z}^1,u_{xy}^1,\dots,
  u^N_{M_1^N x\,M_2^N y\,M_3^N z} ),
\end{equation}
with $M_1^j, M_2^j, M_3^j,$ and $M$ as defined earlier.
Each term in $f$ must be a monomial in jet space variables, either multiplied 
with a constant or variable coefficient.
\begin{defn}
The total derivative operator $\D_x$ in 2\,D is defined as
\begin{equation}
\label{totalderivative2D}
\D_x f =
\frac{\pl f}{\pl x} + \sum_{j=1}^N \sum_{k_1=0}^{M_1^j} \sum_{k_2=0}^{M_2^j}
u_{(k_1 + 1) x \, k_2 y}^j \frac{\pl f}{\pl u_{k_1 x \, k_2 y}^j},
\end{equation}
where $M_1^j$ and $M_2^j$ are the orders of $f$ for component $u^j$ with 
respect to $x$ and $y,$ respectively.
$\D_y$ is defined analogously.
Since $t$ is parameter, $\D_t$ (in 2\,D) is defined in a simpler manner,
\begin{equation}
\label{totalTderivative}
\D_t f = \frac{\pl f}{\pl t} + \sum_{j=1}^N \sum_{k_1=0}^{M_1^j}
\sum_{k_2=0}^{M_2^j}
\frac{\pl f}{\pl u_{k_1 x k_2 y}^j} \D_x^{k_1} \D_y^{k_2} u_t^j.
\end{equation}
\end{defn}
If a total derivative operator were applied by hand to a differential
function, $f(\mf{x},\mf{u}^{(M)}(\mf{x})),$ one would use the product 
and chain rules to complete the computation.
However, formulas like (\ref{totalderivative1D}), (\ref{totalderivative2D}), 
and (\ref{totalTderivative}) are more suitable for symbolic computation.

The Euler operator (also known as the variational derivative) plays a 
fundamental role in the calculus of variations \citep{Olver93}, and 
serves as a key tool in our conservation laws algorithm.
The Euler operator can be defined for any number of independent and 
dependent variables.
For example in 1\,D, the Euler operator is denoted by 
$\mc{L}_{\mf{u} (x)} 
= (\mc{L}_{u^1 (x)}, \mc{L}_{u^2 (x)},\dots , \mc{L}_{u^j (x)}, \dots, 
\mc{L}_{u^N (x)}).$ 
\vskip 0.001pt
\noindent
\begin{defn}
The 1\,D Euler operator for dependent variable $u^j(x)$ is defined as
\begin{eqnarray}
\label{zeroeulerscalarux1D}
\mc{L}_{u^j (x)} f
\!&\!=\!&\!\sum_{k=0}^{M_1^j} (-\D_x)^k \frac{\pl f}{\pl u_{kx}^j}
\nonumber \\
\!&\!=\!&\!
\frac{\pl f}{\pl u^j} - \D_x \frac{\pl f}{\pl u_x^j}
+ \D_x^2 \frac{\pl f}{\pl u_{2x}^j}
- \D_x^3 \frac{\pl f}{\pl u_{3x}^j} + \cdots
+ (-\D_x)^{M_1^j} \frac{\pl f}{\pl u_{M_1^j x}^j},
\end{eqnarray}
$j = 1, \dots, N.$
The 2\,D and 3\,D Euler operators are defined analogously \citep{Olver93}.
For example, the 2\,D Euler operator is 
\begin{equation}
\label{zeroeulerscalarux2D}
\mc{L}_{u^j(x,y)} f
= \sum_{k_1=0}^{M_1^j} \sum_{k_2=0}^{M_2^j} (-\D_x)^{k_1} (-\D_y)^{k_2}
\frac{\pl f}{\pl u_{k_{1}x\,k_{2}y}}, \quad j = 1, \dots, N.
\end{equation}
\end{defn}
The Euler operator allows one to test if differential functions are exact 
which is a key step in the computation of conservation laws.
\begin{defn}
Let $f$ be a differential function of order $M.$
In 1\,D, $f$ is called {\it exact} if $f$ is a total derivative, 
i.e., there exists a differential function 
$F(\mf{x}, \mf{u}^{(M-1)}(\mf{x}))$ such that $f = \D_x F.$
In 2\,D or 3\,D, $f$ is {\it exact} if $f$ is a total divergence,
i.e., there exists a differential vector function 
$\mf{F}(\mathbf{x},\mf{u}^{(M-1)}(\mf{x}))$ such that 
$f = \mathrm{Div}\,\mf{F}.$
\end{defn}
\vskip 0.0001pt
\noindent
\begin{thm}
\label{zeroeulerexact}
A differential function $f$ is exact if and only if 
$\mc{L}_{\mf{u}(\mf{x})} f \equiv \mf{0}.$
Here, $\mf{0}$ is the vector $(0,0,\cdots,0)$ which has $N$ components
matching the number of components of $\mf{u}.$
\vskip 0.001pt
\noindent
\end{thm}
\vskip 0.001pt
\noindent
\begin{pf}
The proof for a general multi-dimensional case is given in, 
e.g., \citet{Poole09}.
\end{pf}
\vskip 0.001pt
\noindent
Next, we turn to the homotopy operator \citep{Anderson04,Olver93}, which 
integrates exact 1\,D differential functions, or inverts the total divergence 
of exact 2\,D or 3\,D differential functions.
Integration routines in CAS have been unreliable when integrating exact 
differential expressions involving unspecified functions.
Often the built-in integration by parts routines fail when arbitrary 
functions appear in the integrand.
The 1\,D homotopy operator offers an attractive alternative since it 
circumvents integration by parts altogether.
\vskip 0.001pt
\noindent
\begin{defn}
\label{oneDhomotopyoperator}
Let $f$ be an exact 1\,D differential function.  
The homotopy operator in 1\,D is defined \citep{Heremanetal07} as
\begin{equation}
\label{1Dhomotopy}
\mc{H}_{\mf{u}(x)} f = \int_0^1
\left( \sum_{j=1}^{N} \mc{I}_{u^j(x)} f \right)
[\lambda {\mf{u}}]\,\frac{d \lambda}{\lambda},
\end{equation}
where $\mf{u} = (u^1, \dots, u^j, \dots, u^N).$
The integrand, $\mc{I}_{u^j(x)} f,$ is defined as
\begin{equation}
\label{1Dintegrand}
\mc{I}_{u^j(x)} f
= \sum_{k=1}^{M_1^j} \left( \sum_{i=0}^{k-1}{u_{ix}^j}
\left( -\D_x \right)^{k-(i+1)} \right) \frac{\pl f}{\pl {u_{kx}^j}},
\end{equation}
where $M_1^j$ is the order of $f$ in dependent variable $u^j$ with respect
to $x.$
The notation $f[\lambda \mf{u}]$ means that in $f$ one replaces
$\mf{u}$ by $\lambda\,\mf{u}, \mf{u}_x$ by $\lambda\,\mf{u}_x,$
and so on for all derivatives of $\mf{u}.$  
$\lambda$ is an auxiliary parameter that traces the homotopic path.
\end{defn}
\vskip 0.001pt
\noindent
Given an exact differential function, the 1\,D homotopy operator
(\ref{1Dhomotopy}) replaces integration by parts (in $x)$ with a sequence of
differentiations followed by a standard integration with respect to $\lambda.$
Indeed, the following theorem states one purpose of the homotopy operator.
\vskip 0.001pt
\noindent
\begin{thm}
\label{homotopy1Dtheorem}
Let $f$ be exact, i.e., $\D_x F = f$ for some differential function 
$F(\mf{x}, \mf{u}^{(M-1)}(\mf{x})).$
Then, $F = \D_x^{-1} f = \mc{H}_{\mf{u}(x)} f.$
\end{thm}
\vskip 0.001pt
\noindent
\begin{pf}
A proof for the 1\,D case in the language of standard calculus is given in
\vskip 1pt
\noindent
\citet{PooleHereman10}.
See \citet{Olver93} for a proof based on the variational complex.
\end{pf}
\vskip 0.001pt
\noindent
The homotopy operator (\ref{1Dhomotopy}) has been a reliable tool for
integrating exact {\it polynomial} differential expressions. 
For applications, see 
\citet{Cheviakov07,Cheviakov10,DeconinckNivala09,Hereman06,Heremanetal07}.
However, the homotopy operator fails to integrate certain classes of
exact rational expressions as discussed in \citet{PooleHereman10}.
Although, the homotopy integrator code in \citet{PooleHereman09software1} 
covers large classes of exact rational functions, we will not consider
rational expressions in this paper. 

CAS often cannot invert the divergences of exact 2\,D and 3\,D differential 
functions, although some capabilities exist in {\it Maple}. 
Again, the homotopy operator is a valuable tool to compute 
$\mathrm{Div}^{-1},$ when it is impossible to do so by hand or by using the
available software tools. 
\vskip 0.001pt
\noindent
\begin{defn}
\label{twoDhomotopyoperator}
The 2\,D homotopy operator is a ``vector'' operator with two components,
\begin{equation}
\label{2Dhomotopyvector}
\left(\mc{H}_{\mf{u}(x,y)}^{(x)} f, \mc{H}_{\mf{u}(x,y)}^{(y)} f \right),
\end{equation}
where
\begin{equation}
\label{2Dhomotopyxandy}
\mc{H}_{\mf{u}(x,y)}^{(x)} f = \int_0^1
\left( \sum_{j=1}^{N} I_{u^j(x,y)}^{(x)} f \right)
[\lambda {\mf{u}}]\,\frac{d \lambda}{\lambda} 
\quad {\rm and} \quad
\mc{H}_{\mf{u}(x,y)}^{(y)} f = \int_0^1
\left( \sum_{j=1}^{N} I_{u^j(x,y)}^{(y)} f \right)
[\lambda {\mf{u}}]\,\frac{d \lambda}{\lambda}.
\end{equation}
The $x$-integrand, $\mc{I}_{u^j(x,y)}^{(x)} f,$ is given by
\begin{equation}
\label{integrand2Dx}
I_{u^j(x,y)}^{(x)} f 
= \sum_{k_1=1}^{M_1^j} \sum_{k_2=0}^{M_2^j}
\left( \sum_{i_1=0}^{k_1-1} \sum_{i_2=0}^{k_2}
B^{(x)}\, u_{i_1 x\,i_2 y}^j \left( -\D_x \right)^{k_1-i_1-1}
\!\left( -\D_y \right)^{k_2-i_2} \right) \frac{\pl f}{\pl u_{k_1 x\,k_2 y}^j},
\end{equation}
with combinatorial coefficient $B^{(x)} = B(i_1,i_2,k_1,k_2),$ where 
\begin{equation}
\label{binomialcoef2D}
B(i_1, i_2, k_1, k_2)  \stackrel{\mathrm{def}}{=} \frac{{i_1 + i_2 \choose i_1}
{k_1 + k_2 - i_1 - i_2 - 1 \choose k_1 - i_1 - 1}}{{k_1 + k_2 \choose k_1}}.
\end{equation}
Similarly, the $y$-integrand, $\mc{I}_{u^j(x,y)}^{(y)} f,$ is defined as
\vspace*{-3mm}
\begin{equation}
\label{integrand2Dy}
I_{u^j(x,y)}^{(y)}f 
= \sum_{k_1=0}^{M_1^j} \sum_{k_2=1}^{M_2^j}
\left( \sum_{i_1=0}^{k_1} \sum_{i_2=0}^{k_2-1}
B^{(y)}\, u_{i_1 x\,i_2 y}^j \left( -\D_x \right)^{k_1-i_1}
\!\left( -\D_y \right)^{k_2-i_2-1} \right)\frac{\pl f}{\pl u_{k_1 x\,k_2 y}^j},
\end{equation}
where $B^{(y)} = B(i_2,i_1,k_2,k_1).$
\end{defn}
\vskip 0.001pt
\noindent
\begin{defn}
\label{threeDhomotopyoperator}
The homotopy operator in 3\,D is a three-component vector operator,
\begin{equation}
\label{3Dhomotopyvector}
\left( \mc{H}_{\mf{u}(x,y,z)}^{(x)} f, \mc{H}_{\mf{u}(x,y,z)}^{(y)} f,
\mc{H}_{\mf{u}(x,y,z)}^{(z)} f \right),
\end{equation}
where the $x$-component is given by
\begin{equation}
\label{3Dhomotopyx}
\mc{H}_{\mf{u}(x,y,z)}^{(x)} f
= \int_{0}^1
\left( \sum_{j=1}^{N} \mc{I}_{u^j(x,y,z)}^{(x)} f \right)
[\lambda {\mf{u}}]\,\frac{d \lambda}{\lambda}.
\end{equation}
The $y$- and $z$-components are defined analogously.
The $x$-integrand is given by
\begin{eqnarray}
\label{integrand3Dx}
I_{u^j(x,y,z)}^{(x)} f\!&=&\!\sum_{k_1=1}^{M_1^j}
\sum_{k_2=0}^{M_2^j} \sum_{k_3=0}^{M_3^j}
\sum_{i_1=0}^{k_1-1} \sum_{i_2=0}^{k_2} \sum_{i_3=0}^{k_3}
\left( B^{(x)}\, u_{i_1 x\,i_2 y\,i_3 z}^j \right.
\nonumber \\
&& \ \ \ \ \ \ \ 
\left. \left( - \D_x \right)^{k_1-i_1-1}
\left( - \D_y \right)^{k_2-i_2} \left( -\D_z \right)^{k_3-i_3} \right)
\frac{\pl f}{\pl u_{k_1 x\,k_2 y\,k_3 z}^j},
\end{eqnarray}
with combinatorial coefficient $B^{(x)} = B(i_1, i_2, i_3, k_1, k_2, k_3)$ 
where
\begin{equation}
\label{binomialcoef3D}
\!\!\!\!\!\!\!\!\!\!
B(i_1, i_2, i_3, k_1, k_2, k_3) \stackrel{\mathrm{def}}{=}
\frac{{i_1 + i_2 + i_3 \choose i_1}\, {i_2 + i_3 \choose i_2}\,
{k_1 + k_2 + k_3 - i_1 - i_2 - i_3 - 1 \choose k_1 - i_1 - 1}\,
{k_2 + k_3 - i_2 - i_3 \choose k_2 - i_2}}
{{k_1 + k_2 + k_3 \choose k_1}\,{k_2 + k_3 \choose k_2}}.
\end{equation}
The integrands $I_{u^j(x,y,z)}^{(y)} f$ and $I_{u^j(x,y,z)}^{(z)} f$ 
are defined analogously.
Based on cyclic permutations, they have combinatorial coefficients
$B^{(y)} = B(i_2,i_3,i_1,k_2,k_3,k_1)$ and
$B^{(z)} = B(i_3,i_1,i_2,k_3,k_1,k_2),$ respectively.
\end{defn}
\vskip 0.001pt
\noindent
Using homotopy operators, $\mathrm{Div}^{-1}$ can be computed based on the 
following theorem.
\vskip 0.001pt
\noindent
\begin{thm}
\label{homotopy2D3Dtheorem}
Let $f$ be exact, i.e.,
$f = \mathrm{Div}\,\mf{F}$ for some $\mf{F}(\mf{x}, \mf{u}^{(M-1)}(\mf{x})).$
Then, in the 2\,D case,
$\mf{F} = \mathrm{Div}^{-1} f =
\left(\mc{H}_{\mf{u}(x,y)}^{(x)} f, \mc{H}_{\mf{u}(x,y)}^{(y)} f \right).$
Analogously, in 3\,D one has 
\newline
\noindent
$\mf{F} = \mathrm{Div}^{-1} f =
\left( \mc{H}_{\mf{u}(x,y,z)}^{(x)} f, \mc{H}_{\mf{u}(x,y,z)}^{(y)} f,
\mc{H}_{\mf{u}(x,y,z)}^{(z)} f \right).$
\end{thm}
\vskip 0.001pt
\noindent
\begin{pf}
A proof for the 2\,D case is given in \citet{Poole09}.
The 3\,D case could be proven with similar arguments.
\end{pf}
\vskip 0.001pt
\noindent
Unfortunately, the outcome of the homotopy operator is not unique.
The homotopy integral in the 1\,D case has a harmless arbitrary constant.
However, in the 2\,D and 3\,D cases there are infinitely many non-trivial
choices for $\mf{F}.$
From vector calculus we know that $\mathrm{Div}\,\mathrm{Curl}\,\mf{K} = 0.$
Thus, the addition of $\mathrm{Curl}\,\mf{K}$ to $\mf{F}$ would not alter
$\mathrm{Div}\,\mf{F}.$
More precisely, for $\mf{K} =  (\D_y \theta, - \D_x \theta)$ in 2\,D, or for
$\mf{K} = (\D_y \eta - \D_z \xi, \D_z \theta - \D_x \eta,
\D_x \xi - \D_y \theta)$ in 3\,D,
$\mathrm{Div}\,\mf{G} = \mathrm{Div}\,(\mf{F} + \mf{K})
= \mathrm{Div}\,\mf{F},$ where $\theta, \eta,$ and $\xi$ are arbitrary
functions.
To obtain a concise result for $\mathrm{Div}^{-1},$ \citet{PooleHereman10} 
developed an algorithm that removes curl terms. 
Furthermore, when $f$ is rational \citep{PooleHereman10}, the homotopy 
operator may fail at the singularities of $f;$ but rational functions are 
not considered in this paper.
%
\section{An Algorithm for Computing a Conservation Law}
\label{secAlgorithm}
To compute a conservation law, the PDE is assumed to be in the form 
given in (\ref{evolutiondefinition}) for a suitable evolution variable.
Adhering to (\ref{genconlaw}), if the evolution variable is $t,$ 
we construct a candidate density.
However, if the evolution variable is $x,$ $y,$ or $z,$ we construct a 
candidate component of the flux corresponding to the evolution variable.
For argument's sake let us assume that the evolution variable is time.

The candidate density is constructed by taking a linear combination 
(with undetermined coefficients) of terms that are invariant under the 
scaling symmetry of the PDE.
The total time derivative of the candidate is computed and evaluated on 
(\ref{evolutiondefinition}), thus removing all time derivatives from the
problem.
The resulting expression must be exact, so we use the Euler operator and 
Theorem~\ref{zeroeulerexact} to derive the linear system that yields the 
undetermined coefficients.
Substituting these coefficients into the candidate leads to a valid density.

Once the density is known the homotopy operator and 
Theorems~\ref{homotopy1Dtheorem} or~\ref{homotopy2D3Dtheorem} are used 
to compute the associated flux, $\mf{J},$ taking advantage of 
(\ref{genconlaw}).

In contrast to other algorithms which attempt to compute the components 
of $\mf{P}$ in (\ref{divconlaw}) all at once, our algorithm computes the 
density first, followed by the flux.
Although restricted to polynomial conservation laws, our constructive method 
leads to short densities (which are free of divergences and  
divergence-equivalent terms) and fluxes in which all curl terms are 
automatically removed.
\noindent
\begin{defn}
A term or expression $f$ is a {\it divergence} if there exists a vector
{\bf F} such that $f = \mathrm{Div}\,\mathbf{F}.$
In the 1\,D case, $f$ is a total derivative if there exists a function 
$F$ such that $f = \D_x F.$
Note that $\D_x f$ is essentially a one-dimensional divergence.
So, from here onwards, the term ``divergence'' will also cover the 
``total derivative" case.
Two or more terms are {\it divergence-equivalent} when a linear combination 
of the terms is a divergence.
\end{defn}
\noindent
To illustrate the subtleties of the algorithm we intersperse the steps of the 
algorithm with {\it two} examples, viz., the ZK and KP equations.
%
\subsection{Computing the Scaling Symmetry}
\label{secScalingSymmetry}
A PDE has a unique set of Lie-point symmetries which may include translations,
rotations, dilations, Galilean boosts, and other symmetries 
\citep{Blumanetal10}. 
The application of such symmetries allows one to generate new solutions 
from known solutions.
We will use only one type of Lie-point symmetry, namely, the scaling or 
dilation symmetry, to formulate a ``candidate density.''

Let us assume that a PDE has a scaling symmetry.
For example, the ZK equation (\ref{ZKequation2D}) is invariant under the 
scaling symmetry
\begin{equation}
\label{ZKscalingsymmetry}
(x,y,t,u) \rightarrow ( \lambda^{-1}x, \lambda^{-1}y, \lambda^{-3}t, 
\lambda^2 u),
\end{equation}
where $\lambda$ is an arbitrary scaling parameter, not to be confused with 
$\lambda$ in Definitions~\ref{oneDhomotopyoperator} 
through~\ref{threeDhomotopyoperator}.
\vskip 8pt
\noindent
{\bf Step 1-ZK} (Computing the scaling symmetry).
To compute (\ref{ZKscalingsymmetry}) with linear algebra, assume that 
(\ref{ZKequation2D}) for $u(x,y,t)$ scales uniformly under
\begin{equation}
\label{scalesymmetryunknown}
(x,y,t,u) \rightarrow (X,Y,T,U) \equiv ( \lambda^{a}x, \lambda^{b}y,
\lambda^{c}t, \lambda^{d}u),
\end{equation}
where $U(X,Y,T)$ and $a,$ $b,$ $c,$ and $d$ are undetermined (rational) 
exponents. 
We assume that the parameters $\alpha$ and $\beta$ do not scale.
By the chain rule, (\ref{ZKequation2D}) transforms into
\begin{eqnarray}
\label{ZKwithunknownscaledsym}
u_t \!\!\! &+& \!\!\! \alpha u u_x + \beta u_{3x} + \beta u_{x2y}
\nonumber
\\ \!\!\! &=& \!\!\! \lambda^{c-d} \left( U_T + \alpha \lambda^{a-c-d}U U_X
+ \beta \lambda^{3a-c}U_{3X} + \beta \lambda^{a+2b-c} U_{X2Y} \right) = 0.
\end{eqnarray}
If $a - c - d = 3 a - c = a + 2b - c = 0,$ we have (\ref{ZKequation2D}) for
$U(X,Y,T)$ up to the scaling factor $\lambda^{c-d}.$
Setting $a = -1,$ we find $b = -1, c = -3,$ and $d = 2,$ corresponding to 
(\ref{ZKscalingsymmetry}).
\vskip 8pt
\noindent 
{\bf Step 1-KP} (Computing the scaling symmetry).
The scaling symmetry for the KP equation will be computed similarly.
Assume that (\ref{KPevolutionform}) scales uniformly under
\begin{equation}
\label{scalesymmetryunknown2}
(x,y,t,u,v) \rightarrow (X,Y,T,U,V) \equiv ( \lambda^{a}x, \lambda^{b}y,
\lambda^{c}t, \lambda^{d}u, \lambda^{e}v),
\end{equation}
with unknown rational exponents $a$ through $e.$
Applying the chain rule to get (\ref{KPevolutionform}) expressed in the 
variables $(X,Y,T,U,V)$ yields
\begin{eqnarray}
\label{KPKwithunknownscaledsym}
\!\!\!\!\!\!\!\!\! 
u_y \!\!\! & - & \!\!\! v 
= \lambda^{b-d}(U_Y - \lambda^{d-b-e}V) = 0,
\nonumber \\ 
\!\!\!\!\!\!\!\!\! 
v_y \!\!\! & + & \!\!\!\sigma^2 (u_{tx} + u_x^2 + u u_{2x} + u_{4x})
\nonumber \\
\!\!\!\!\!\!\!\!\! 
\!\!\!\!\! & = & \!\!\! 
\lambda^{b-e} \left( V_Y + \sigma^2 \left( \lambda^{a-b+c-d+e} U_{TX} 
+ \alpha \lambda^{2a-b-2d+e} (U_X^2 + U U_{2X}) 
+ \lambda^{4a-b-d+e} U_{4X} \right) \right) = 0.
\end{eqnarray}
By setting  
$d - b - e = a - b + c - d + e = 2a - b - 2d + e = 4a - b- d + e = 0,$ 
(\ref{KPKwithunknownscaledsym}) becomes a scaled version of 
(\ref{KPevolutionform}) in the new variables $U(X,Y,T)$ and $V(X,Y,T).$
Setting $a = -1$ yields $b = -2,$ $c = -3,$ $d = 2,$ and $e = 4.$
Hence,
\begin{equation}
\label{KPscalingsymmetry}
(x,y,t,u,v) \rightarrow ( \lambda^{-1}x, \lambda^{-2}y, \lambda^{-3}t,
\lambda^2 u, \lambda^4 v)
\end{equation}
is a scaling symmetry of (\ref{KPevolutionform}).
%
\subsection{Constructing a Candidate Component}
\label{secCandidateComponent}
Conservation law (\ref{divconlaw}) must hold on solutions of the PDE.
Therefore, we search for polynomial conservation laws that obey the 
scaling symmetry of the PDE.
Indeed, we have yet to find a polynomial conservation law that does not 
adhere to the scaling symmetry.

Based on the scaling symmetry of the PDE, we choose a scaling factor for one 
of the components of $\mf{P}$ in (\ref{divconlaw}).
The selected scaling factor will be called the {\it rank} $(R)$ of that 
component.
Then, we construct a candidate for that component as a linear combination of 
monomial terms (all of rank $R$) with undetermined coefficients.
By dynamically removing divergence terms and divergence-equivalent terms that
candidate is short and of low order. 
\vskip 8pt
\noindent 
{\bf Step 2-ZK} (Building the candidate component).
Since the ZK equation (\ref{ZKequation2D}) has $t$ as evolution variable, 
we will compute the density $\rho$ of (\ref{genconlaw}) of a fixed rank,
for example, $R = 6.$
\vskip 5pt
\noindent 
{\bf (a)} 
Construct a list, $\mc{P},$ of differential terms containing all powers 
of dependent variables and products of dependent variables that have 
rank 6 or less.
By (\ref{ZKscalingsymmetry}), $u$ has a scaling factor of 2, so $u^3$ 
scales to rank 6 and $u^2$ has rank 4.
This leads to $\mc{P} = \{ u^3, u^2, u \}.$
\vskip 5pt
\noindent {\bf (b)} 
Bring all of the terms in $\mc{P}$ up to rank 6 and put them into a new list,
$\mc{Q}.$
This is done by applying the total derivative operators with respect to the 
space variables.
Taking the terms in $\mc{P}, u^3$ has rank 6 and is placed directly into
$\mc{Q}.$ 
The term $u^2$ has rank 4 and can be brought up to rank six in three ways: 
either by applying $\D_x$ twice, by applying $\D_y$ twice, or by applying 
each of $\D_x$ and $\D_y$ once, since both $\D_x$ and$\D_y$ have 
scaling factors of 1.
All three possibilities are considered and the resulting terms are put into 
$\mc{Q}.$
Similarly, the term $u$ can be brought up to rank 6 in five ways,
and all results are placed into $\mc{Q}.$
Doing so, 
\begin{equation}
\label{ZKrank6Qlist}
\mc{Q} = \{u^3, u_x^2, u u_{2x}, u_y^2, u u_{2y}, u_x u_y, u u_{xy}, u_{4x},
u_{3xy}, u_{2x2y}, u_{x3y}, u_{4y}\},
\end{equation}
in which all monomials are now of rank 6.
\vskip 5pt
\noindent
{\bf (c)} 
With the goal of constructing a nontrivial density with the least number of 
terms, remove all terms that are {\it divergences} or are 
{\it divergence-equivalent} to other terms in $\mc{Q}.$
This can be done algorithmically by applying the Euler operator 
(\ref{zeroeulerscalarux2D}) to each term in (\ref{ZKrank6Qlist}), yielding
\begin{equation}
\label{ZKzeroEulerofQ}
\mc{L}_{u(x,y)} \mc{Q} = \{3u^2, -2 u_{2x}, 2 u_{2x}, -2 u_{2y},
2 u_{2y}, -2 u_{xy}, 2 u_{xy}, 0, 0, 0, 0, 0\}.
\end{equation}
By Theorem \ref{zeroeulerexact}, divergences are terms corresponding to $0$ 
in (\ref{ZKzeroEulerofQ}).
Hence, $u_{4x},$ $u_{3xy},$ $u_{2x2y},$ $u_{x3y},$ and $u_{4y}$ are 
divergences and can be removed from $\mc{Q}.$
Next, all divergence-equivalent terms will be removed.
Following \citet{HeremanAll05}, form a linear combination of the terms that 
remained in (\ref{ZKzeroEulerofQ}) with undetermined coefficients $p_i,$ 
gather like terms, and set it identically equal to zero, 
\begin{equation}
\label{linearindependence}
3 p_1 u^2 + 2 (p_3 - p_2) u_{2x} + 2 (p_5 - p_4) u_{2y}
+ 2 (p_7 - p_6) u_{xy} = 0.
\end{equation}
Hence, $p_1 = 0,$ $p_2 = p_3,$ $p_4 = p_5,$ and $p_6 = p_7.$
Terms with coefficients $p_3,$ $p_5,$ and $p_7$ are divergence-equivalent 
to the terms with coefficients $p_2.$ $p_4,$ and $p_6,$ respectively.
For each divergence-equivalent pair, the terms of highest order are removed 
from $\mc{Q}$ in (\ref{ZKrank6Qlist}).
After all divergences and divergence-equivalent terms are removed,
$\mc{Q} = \{ u^3, u_x^2, u_y^2, u_x u_y \}.$
\vskip 3pt
\noindent 
{\bf (d)} 
A candidate density is obtained by forming a linear combination of the 
remaining terms in $\mc{Q}$ using undetermined coefficients $c_i.$ 
Thus, the candidate density of rank 6 for (\ref{ZKequation2D}) is
\begin{equation}
\label{ZKcandidaterank6}
\rho = c_1 u^3 + c_2 u_x^2 + c_3 u_y^2 + c_4 u_x u_y.
\end{equation}
\vskip 10pt
Now, we turn to the KP equation (\ref{KPequation}).
The conservation laws for the KP equation, 
(\ref{KPgenconlaw1}) and (\ref{KPgenconlaw2}), involve an arbitrary 
functional coefficient $f(t).$
The scaling factor for $f(t)$ depends on the degree if $f(t)$ is polynomial;
whereas there is no scaling factor if $f(t)$ is non-polynomial.
In general, working with undetermined functional (instead of constant) 
coefficients $f(x,y,z,t)$ would require a sophisticated solver for PDEs 
for $f$ (see \citet{Wolf02}).
Therefore, we can {\it not automatically} compute (\ref{KPgenconlaw1}) and 
(\ref{KPgenconlaw2}) with our method. 
However, our algorithm can find conservation laws with 
explicit variable coefficients, e.g., $t x^2, t x y,$ etc., as long as the 
degree is specified.
Allowing such coefficients causes the candidate component to have a 
negative rank.
By computing several conservation laws with explicit variable coefficients 
it is possible (by pattern matching) to guess and subsequently test the form 
of a conservation law with arbitrary functional coefficients. 
\vskip 8pt
\noindent 
{\bf Step 2-KP} (Building a candidate $y$-component).
When the KP equation is replaced by (\ref{KPevolutionform}), the evolution 
variable is $y.$ 
Thus, we will compute a candidate for the $y$-component of the flux, $J^y,$ 
in (\ref{genconlaw}).
The $y$-component will have rank equal to $-3.$ 
The negative rank occurs since differential terms for the component are 
multiplied by $c_i\, t^m x^n y^p,$ which, by (\ref{KPscalingsymmetry}), 
scales with $\lambda^{-3m} \lambda^{-n}\lambda^{-2p} = \lambda^{-(3m+n+2p)},$
where $m,$ $n,$ and $p$ are positive integers.
The total degree of the variable coefficient $t^m x^n y^p,$ is restricted 
to $0 \le m+n+p \le 3.$
\vskip 5pt
\noindent 
{\bf (a)} 
As shown in Table~\ref{KPcandidatetable}, construct two lists,
one with all possible coefficients $t^m x^n y^p$ up to degree 3 and the other 
with differential terms, organized so that the combined rank equals $-3.$
The rank of each term is computed using the scaling factors from 
(\ref{KPscalingsymmetry}).
For example, $t$ and $x$ have scaling factors of $-3$ and $-1,$ 
respectively, so $tx^2$ has rank $-5.$  
Variable $u$ has scaling factor 2, so $t^2 x u$ has rank $-3.$
Since we are computing the $y$-component of $\mf{J},$ the differential terms 
contain only derivatives with respect to $x$ and $t.$
\noindent
\begin{table}[h!]
\begin{center}
\begin{tabular}{|r|l||r|l||r|} \hline
\multicolumn{2}{|c||}{\rule[-2mm]{0mm}{7mm}
Factors of Type $t^m x^n y^p$} &
\multicolumn{2}{|c||}{\rule[-2mm]{0mm}{7mm} Differential Terms} &
\multicolumn{1}{|c|}{\rule[-2mm]{0mm}{7mm} Product} 
\\ \hline Rank & Coefficient & Rank & Term & Rank 
\\ \hline \hline $-5$ & $t x^2,$ $x y^2,$ $t y$ & 2 & $u$ & -3
\\ \hline $-6$ & $y^3,$ $t x y,$ $t^2$ & 3 & $u_x$ & -3
\\ \hline $-7$ & $t^2 x,$ $t y^2$ & 4 & $u^2,$ $u_{2x},$ $v$ & -3
\\ \hline $-8$ & $t^2 y$ & 5 & $u u_x,$ $u_t,$ $u_{3x},$ $v_x$ & -3
\\ \hline $-9$ & $t^3$ & 6 & $u^3,$ $uv,$ $u_x^2,$ $u_{tx},$ $u u_{2x},$
$u_{4x},$ $v_{2x}$ & -3
\\ \hline
\end{tabular}
\end{center}
\caption{
Factors $t^m x^n y^p$ of degree 3 are paired with differential terms so that 
their products have ranks $-3.$ }
\label{KPcandidatetable}
\end{table}
\vskip 0.0001pt
\noindent
{\bf (b)} 
Combine the terms in Table~\ref{KPcandidatetable} to create a list 
of all possible terms with rank $-3,$
\begin{eqnarray}
\label{KPrankneg3Qterms}
\mc{Q} \!\!\! &=& \!\!\! \{ t x^2 u, x y^2 u, t y u, y^3 u_x, t x y u_x,
t^2 u_x, t^2 x u^2, t y^2 u^2, t^2 x u_{2x}, t y^2 u_{2x}, t^2 x v, t y^2 v
\nonumber \\ 
&& \!\!\!\!\!\!\!\!\!
t^2 y u u_x, t^2 y u_t, t^2 y u_{3x}, t^2 y v_x, t^3 u^3, t^3 u v,
t^3 u_x^2, t^3 u_{tx}, t^3 u u_{2x}, t^3 u_{4x}, t^3 v_{2x} \}.
\end{eqnarray}
\noindent
{\bf (c)} Remove all divergences and divergence-equivalent terms.
Apply the Euler operator to each term in (\ref{KPrankneg3Qterms}). 
Next, linearly combine the resulting terms to get
\begin{eqnarray}
\label{KPrankneg3Eulerimage}
&& \!\!\!\!\!\!\!\!\!\!\!\!\!\!\!\!\!
p_1 \left(\! \begin{array}{c} t x^2 \\ 0 \end{array} \!\!\right)
\!+\!p_2 \left(\!\! \begin{array}{c} x y^2 \\ 0 \end{array} \!\!\right)
\!+\!p_3 \left(\!\! \begin{array}{c} t y \\ 0 \end{array} \!\!\right)
\!-\!p_5 \left(\!\! \begin{array}{c} t y \\ 0 \end{array} \!\!\right)
\!+\!p_7 \left(\!\! \begin{array}{c} 2 t^2 x u \\ 0 \end{array} \!\!\right)
\!+\!p_8 \left(\!\! \begin{array}{c} 2 t y^2 u \\ 0 \end{array} \!\!\right)
\!+\!p_{11} \left(\!\! \begin{array}{c} 0 \\ t^2 x \end{array} \!\!\right)
\nonumber \\ 
&& \!\!\!\!\!\!\!\!\!\!\!\!\!\!\!\!\! 
{} \!+\!p_{12}\left(\!\! \begin{array}{c} 0 \\ t y^2 \end{array} \!\!\right)
\!-\!p_{14}\left(\!\! \begin{array}{c} 2 t y \\ 0 \end{array} \!\!\right)
\!+\!p_{17}\left(\!\! \begin{array}{c} 3 t^3 u^2 \\ 0 \end{array} \!\!\right)
\!+\!p_{18}\left(\!\! \begin{array}{c} t^3 v \\ t^3 u \end{array} \!\!\right)
\!-\!p_{19}\left(\!\!\! 
\begin{array}{c} 2 t^3 u_{2x}\\ 0 \end{array}\!\!\!\right)
\!+\!p_{21}\left(\!\!\! 
\begin{array}{c} 2 t^3 u_{2x}\\ 0 \end{array}\!\!\!\right) 
\!=\! 0,
\end{eqnarray}
where the subscript of the undetermined coefficient, $p_i,$ corresponds to 
the $i$th term in $\mc{Q}.$ 
Missing $p_i$ correspond to terms that are divergences.
Gather like terms, set their coefficients equal to zero, and solve the 
resulting linear system for the $p_i,$ to get
$p_1 = p_2 = p_7 = p_8 = p_{11} = p_{12} = p_{17} = p_{18} = 0,$
$p_3 = p_5 + 2p_{14},$ and $p_{19} = p_{21}.$
Thus, both terms with coefficients $p_5$ and $p_{14}$ are
divergence-equivalent to the term with coefficient $p_3.$ 
Likewise, the term with coefficient $p_{21}$ is divergence-equivalent to the 
term with coefficient $p_{19}.$
For each divergence-equivalent pair, the terms with the highest order are 
removed from (\ref{KPrankneg3Qterms}).
After removal of divergences and divergence-equivalent terms
\begin{equation}
\label{KPfinallistQ}
\mc{Q} = \{ t x^2 u, x y^2 u, t y u, t^2 x u^2, t y^2 u^2,
t^2 x v, t y^2 v, t^3 u^3, t^3 uv, t^3 u_x^2 \}.
\end{equation}
\noindent
{\bf (d)} A linear combination of the terms in (\ref{KPfinallistQ}) with 
undetermined coefficients $c_i$ yields the candidate (of rank $-3)$ 
for the $y$-component of the flux, i.e.,
\begin{eqnarray}
\label{KPcandidaterankneg3}
J^y \!\! &=& \!\!  c_1 t x^2 u + c_2 x y^2 u + c_3 t y u + c_4 t^2 x u^2
+ c_5 t y^2 u^2 + c_6 t^2 x v
\nonumber \\ 
&& \!\! {} + c_7 t y^2 v + c_8 t^3 u^3 + c_9 t^3 u v + c_{10} t^3 u_x^2.
\end{eqnarray}
%
\subsection{Evaluating the Undetermined Coefficients}
\label{secActualComponent}
All, part, or none of the candidate density (\ref{ZKcandidaterank6}) may
be an actual density for the ZK equation.
It is also possible that the candidate is a linear combination of two or
more independent densities, yielding independent conservation laws.  
The true nature of the density will be revealed by computing the 
undetermined coefficients.
By (\ref{genconlaw}), $\D_t \rho = -\mathrm{Div}\,(J^x,J^y),$
so $\D_t \rho$ must be a divergence with respect to the space 
variables $x$ and $y.$
Using Theorem~\ref{zeroeulerexact}, an algorithm for computing the 
undetermined coefficients readily follows.
\vskip 8pt
\noindent 
{\bf Step 3-ZK} (Computing the undetermined coefficients).
To compute the undetermined coefficients, we form a system of linear 
equations for these coefficients.
As part of the solution process, we also generate compatibility conditions 
for the constant parameters in the PDE, if present.
\vskip 5pt
\noindent {\bf (a)} Compute the total derivative with respect to $t$ of
(\ref{ZKcandidaterank6}),
\begin{equation}
\label{DtZK}
\D_t \rho = 3 c_1 u^2 u_t + 2 c_2 u_x u_{tx}
+ 2 c_3 u_y u_{ty} + c_4 (u_{tx} u_y + u_x u_{ty}).
\end{equation}
Let $E = -\D_t \rho$ after $u_t$ and $u_{tx}$ have been replaced using 
(\ref{ZKequation2D}).
This yields
\begin{eqnarray}
\label{ZKrank6dtrho}
E \!\!\! 
&=& \!\!\! 3 c_1 u^2 (\alpha u u_x + \beta (u_{3x} + u_{x2y})) 
+ 2 c_2 u_x (\alpha u u_x + \beta (u_{3x} + u_{x2y}))_x
\nonumber
\\ 
&& \!\!\! {} + 2 c_3 u_y (\alpha u u_x + \beta (u_{3x} + u_{x2y}))_y
+ c_4 (u_y (\alpha u u_x + \beta (u_{3x} + u_{x2y}))_x
\nonumber
\\ && 
\!\!\! {} + u_x (\alpha u u_x + \beta (u_{3x} + u_{x2y}))_y).
\end{eqnarray}
\noindent
{\bf (b)} By (\ref{genconlaw}), $E = \mathrm{Div}\,(J^x,J^y).$ 
Therefore, by Theorem \ref{zeroeulerexact}, 
$\mc{L}_{u(x,y)}\ E \equiv \mf{0}.$
Apply the Euler operator to (\ref{ZKrank6dtrho}), gather like terms, and set 
the result identically equal to zero:
\begin{eqnarray}
\label{KPEuleronE}
0 \equiv \mathcal{L}_{u(x,y)}\ E 
\!\! &=& \!\! 
-2 \big( (3 c_1  \beta  + c_3 \alpha) u_x u_{2y} 
+ 2 (3 c_1 \beta + c_3 \alpha) u_y u_{xy}
\nonumber \\ 
&& {} \!\!+ 2 c_4 \alpha u_x u_{xy} + c_4 \alpha u_y u_{2x} 
+ 3 (3 c_1 \beta + c_2 \alpha) u_x u_{2x} \big).
\end{eqnarray}
\noindent
{\bf (c)} Form a {\it linear} system of equations for the undetermined 
coefficients $c_i$ by setting each coefficient equal to zero, 
thus satisfying (\ref{KPEuleronE}).
After eliminating duplicate equations, the system is
\begin{equation}
\label{ZKrank6coefsystem}
3 c_1 \beta + c_3 \alpha = 0, \qquad c_4 \alpha = 0,
\qquad 3 c_1 \beta + c_2 \alpha = 0.
\end{equation}
\noindent
{\bf (d)} Check for possible compatibility conditions on the parameters 
$\alpha$ and $\beta$ in (\ref{ZKrank6coefsystem}).
This is done by setting each $c_i \!=\! 1,$ one at a time, and 
algebraically eliminating the other undetermined coefficients.
Consult \citet{GoktasHeremanJSC97} for details about searching for 
compatibility conditions.
System (\ref{ZKrank6coefsystem}) is compatible for all nonzero $\alpha$ and 
$\beta.$
\vskip 5pt
\noindent 
{\bf (e)} Solve (\ref{ZKrank6coefsystem}), taking into account the 
compatibility conditions (if applicable).
Here, 
\begin{equation}
\label{ZKrank6coefs}
c_2 = c_3 = -3 \tfrac{\beta}{\alpha} c_1, \qquad c_4 = 0,
\end{equation}
where $c_1$ is arbitrary. 
We set $c_1 = 1$ so that the density is normalized on the highest degree term,
yielding
\begin{equation}
\label{ZKrank6finaldensity}
\rho = u^3 -3 \tfrac{\beta}{\alpha} (u_x^2 + u_y^2).
\end{equation}
\vskip 6pt
\noindent
{\bf Step 3-KP} (Computing the undetermined coefficients).  
The procedure to find the undetermined coefficients in the KP case is similar 
to that of the ZK case.
\vskip 5pt
\noindent 
{\bf (a)} 
Starting from (\ref{KPcandidaterankneg3}), compute
\begin{eqnarray}
\label{KPdycandidate}
\D_y J^y \!\!\! 
&=& \!\!\! (c_1 x + 2 c_4 t u) t x u_y + c_2 x y (2 u + y u_y)
+ (c_3 t + 2 c_5 t y u) (u + y u_y)
\nonumber
\\ && 
\!\!{} + c_6 t^2 x v_y + c_7 t y (2 v + y v_y) + 3 c_8 t^3 u^2 u_y
+ c_9 t^3 ( u_y v + u v_y) + 2 c_{10} t^3 u_x u_{xy}, 
\end{eqnarray}
and replace $u_y$ and $v_y$ and their differential consequences using 
(\ref{KPevolutionform}). 
Thus,
\begin{eqnarray}
\label{KPdycandidatereplaced}
E = -\D_y J^y \!\!\! &=& \!\!\!
- (c_1 x + 2 c_4 t u) t x v - c_2 x y (2 u + y v) 
- (c_3 t + 2 c_5 t y u) (u + y v)
\nonumber
\\ 
&& \!\!\! {} + \sigma^2 (c_6 t^2 x +c_7 t y^2 + c_9 t^3 u)
(u_{tx} + \alpha u_x^2 + \alpha u u_{2x} + u_{4x}) -2 c_7 t y v
\nonumber \\ 
&& \!\!\! {} - 3 c_8 t^3 u^2 v - c_9 t^3 v^2 - 2 c_{10} t^3 u_x v_x.
\end{eqnarray}
\noindent
{\bf (b)} 
Apply the Euler operator to (\ref{KPdycandidatereplaced}) and set the result 
identically equal to zero.
This yields
\begin{eqnarray}
\label{KPdycandidateEuler}
(0,\, 0) \!\!&\!=\!&\!\! \mf{0} \!\equiv\! \mc{L}_{\mf{u}(t,x)} \ E 
\!=\! \Big(\mc{L}_{u(t,x)}\ E,\, \mc{L}_{v(t,x)}\ E  \Big) 
\nonumber \\ 
\!\!&\!=\!&\!\!
- \Big(2 c_2 x y + (c_3 - 2 \sigma^2 c_6) t + 2 c_4 t^2 x v
+ 2 c_5 t y (2 u +  y v) + 6 c_8 t^3 u v
\nonumber
\\ && {} \!\!- 2 \sigma^2 c_9 t^2 (\tfrac{3}{2} u_x + t u_{tx} + \alpha t u_x^2
+ \alpha u u_{2x} + t u_{4x}) - 2 c_{10} t^3 v_{2x}, \, 
c_1 t x^2 + c_2 x y^2
\nonumber
\\ && {} \!\!+ (c_3 + 2 c_7) t y + 2 c_4 t^2 x u + 2 c_5 t y^2 u 
+ 3 c_8 t^3 u^2 + 2 c_9 t^3 v - 2 c_{10} t^3 u_{2x} \Big).
\end{eqnarray}
\noindent
{\bf (c)} 
Form a linear system for the undetermined coefficients $c_i.$
After duplicate equations and common factors have been removed, one gets
\begin{equation}
c_1 = 0, \, c_2 = 0, \, c_3 - 2 \sigma^2 c_6 = 0, \, c_3 + 2 c_7 = 0, \, 
c_4 = 0, \, c_5 = 0, \, c_8 = 0, \, c_9 = 0, \, c_{10} = 0.
\end{equation}
\noindent
{\bf (d)} 
Compute potential compatibility conditions on the parameters $\alpha$ and 
$\sigma.$
Again, the system is compatible for all nonzero values of $\alpha$ and 
$\sigma.$
\vskip 5pt
\noindent
{\bf (e)} 
Use $\sigma^2 = \pm 1$ and solve the linear system, yielding 
\begin{equation}
\label{KPrankneg3coefvalues}
c_1 = c_2 = c_4 = c_5 = c_8 = c_9 = c_{10} = 0,\;\;\;
c_6 = \tfrac{1}{2} \sigma^2 c_3, \;\;\; c_7 = -\tfrac{1}{2} c_3.
\end{equation}
Set $c_3 = -2$ (to normalize the density) and substitute 
the result into (\ref{KPcandidaterankneg3}), to obtain
\begin{equation}
\label{KPrankneg3actyflux}
J^y = -t ( 2 y u + (\sigma^2 t x - y^2) v),
\end{equation}
which matches $J^y$ in (\ref{KPgenconlaw1}) if $f(t) = t^2$ and $v = u_y.$
%
\subsection{Completing the Conservation Law}
\label{secConservationLaw}
With the density (or a component of the flux at hand), the remaining 
components of the conservation law can be computed with the homotopy 
operator using Theorem~\ref{homotopy1Dtheorem} or~\ref{homotopy2D3Dtheorem}.
\vskip 8pt
\noindent
{\bf Step 4-ZK} (Computing the flux, $\mf{J}$).
Again, by the continuity equation (\ref{genconlaw}),
$\mathrm{Div}\,\mf{J} = \mathrm{Div}\,(J^x,J^y) = -\D_t \rho = E.$  
Therefore, compute $\mathrm{Div}^{-1}\,E,$ where the divergence is with 
respect to $x$ and $y.$
After substitution of (\ref{ZKrank6coefs}) with $c_1 = 1$ into 
(\ref{ZKrank6dtrho}),
\begin{eqnarray}
\label{EforZK}
E \!\!\! &=& \!\!\! 
3 u^2 (\alpha u u_x  + \beta u_{3x} + \beta u_{x2y} ) 
-6 \tfrac{\beta}{\alpha} u_x (\alpha u u_x  + \beta u_{3x} + \beta u_{x2y})_x
\nonumber \\ && \!\!\! {} 
-6 \tfrac{\beta}{\alpha} u_y (\alpha u u_x  + \beta u_{3x} + \beta u_{x2y} )_y.
\end{eqnarray}
Apply the 2\,D homotopy operator from Theorem~\ref{homotopy2D3Dtheorem}.
Compute the integrands (\ref{integrand2Dx}) and (\ref{integrand2Dy}):
\begin{eqnarray}
\label{integrandxZK}
I_{u(x,y)}^{(x)} E
\!\!\! &=& \!\!\!
3 \alpha u^4 + \beta \left( 9 u^2 (u_{2x} + \tfrac{2}{3} u_{2y})
- 6  u (3 u_x^2 + u_y^2) \right)
+ \tfrac{\beta^2}{\alpha} \left( 6 u_{2x}^2
+ 5 u_{xy}^2 + \tfrac{3}{2} u_{2y}^2 \right.
\nonumber \\
&& \!\!\! {} 
\left. + \tfrac{3}{2} u ( u_{2x2y} + u_{4y} )  
- u_x (12 u_{3x} + 7 u_{x2y})
- u_y (3 u_{3y} + 8 u_{2xy}) + \tfrac{5}{2} u_{2x} u_{2y} \right),
\\ 
\label{integrandyZK}
I_{u(x,y)}^{(y)} E \!\!\! &=& \!\!\! 3 \beta  u ( u u_{xy} - 4 u_x u_y)
- \tfrac{1}{2} \tfrac{\beta^2}{\alpha} \left( 3 u ( u_{3xy} + u_{x3y})
+ u_x (13 u_{2xy} + 3 u_{3y}) \right.
\nonumber \\
&& \!\!\! 
\left. {} + 5 u_y (u_{3x} + 3 u_{x2y}) - 9 u_{xy} ( u_{2x} + u_{2y}) \right),
\end{eqnarray}
respectively. 
Use (\ref{2Dhomotopyxandy}), to compute
$\mf{\hat{J}} = \left( \mc{H}_{\mf{u}(x,y)}^{(x)} E, 
\mc{H}_{\mf{u}(x,y)}^{(y)} E \right)$ where
\begin{eqnarray}
\label{HxtoEZK}
\mc{H}_{\mf{u}(x,y)}^{(x)} E
\!\!\! &=& \!\!\! 
\int_0^1 \!\!\left( \mc{I}_{u(x,y)}^{(x)} E \right)
[\lambda \mf{u}]\,\frac{d \lambda}{\lambda}
\nonumber \\
\!\!\! &=& \!\!\! 
\tfrac{3}{4} \alpha u^4 + \beta \left( 3 u^2 (u_{2x} + \tfrac{2}{3} u_{2y})
- 2  u (3 u_x^2 + u_y^2) \right) + \tfrac{\beta^2}{\alpha} \left( 3 u_{2x}^2 
+ \tfrac{5}{2} u_{xy}^2 + \tfrac{3}{4} u_{2y}^2 \right.
\nonumber \\
&& \!\!\! 
\left. {} 
+ \tfrac{3}{4} u (u_{2x2y} + u_{4y}) - u_x (6 u_{3x} + \tfrac{7}{2} u_{x2y}) 
- u_y (\tfrac{3}{2} u_{3y} + 4 u_{2xy}) + \tfrac{5}{4} u_{2x} u_{2y} \right), 
\\
\label{HytoEZK}
\mc{H}_{\mf{u}(x,y)}^{(y)} E
&\!\!=\!\!& \int_0^1 \!\!\left( \mc{I}_{u(x,y)}^{(x)} E \right)
[\lambda \mf{u}]\,\frac{d \lambda}{\lambda}
\nonumber \\
\!\!\! &=& \!\!\!  
\beta  u ( u u_{xy} - 4 u_x u_y) - \tfrac{1}{4} \tfrac{\beta^2}{\alpha} 
\left( 3 u ( u_{3xy} + u_{x3y}) + u_x (13 u_{2xy} + 3 u_{3y}) \right.
\nonumber \\
&& \!\!\! \left. {} + 5 u_y (u_{3x} + 3 u_{x2y}) 
- 9 u_{xy} ( u_{2x} + u_{2y}) \right).
\end{eqnarray}
Notice that $\mf{\hat{J}}$ has a curl term, 
$\mf{K} = (\D_y \theta, -\D_x \theta),$ with
\begin{equation}
\label{thetaZK}
\theta = 
2 \beta u^2 u_y + 
\tfrac{1}{4} \tfrac{\beta^2}{\alpha} \Big( 3 u ( u_{2xy} + u_{3y} )
+ 5 (2 u_x u_{xy} + 3 u_y u_{2y} + u_{2x} u_y) \Big).
\end{equation}
Therefore, compute $\mf{\hat{J}} - \mf{K}$ to obtain
\begin{eqnarray}
\label{xfluxKZ}
J^x \!\!\! &=& \!\!\! 
3 \Big( u^2 (\tfrac{1}{4} \alpha u^2 + \beta u_{2x} ) 
- 2 \beta u (u_x^2 + u_y^2) 
+ \tfrac{\beta^2}{\alpha} (u_{2x}^2 - u_{2y}^2) 
\nonumber
\\ \!\!\!\!\!\!\!\!\!\!\!\!\!\!\!\! && \hspace*{5mm} {}
- 2 \tfrac{\beta^2}{\alpha} (u_x (u_{3x} + u_{x 2y})
+ u_y (u_{2x y} + u_{3y})) \Big),
\\ \label{yfluxKZ}
J^y \!\!\! &=& \!\!\! 3 \beta \Big( 
u^2 u_{xy} + 2 \tfrac{\beta}{\alpha} u_{xy} (u_{2x} + u_{2y}) \Big),
\end{eqnarray}
which match the components in (\ref{ZK2Dconlaw3}).
\vskip 8pt
\noindent
{\bf Step 4-KP} (Computing the density and the $x$-component of the flux).
For the KP example, $(\rho, J^x)$ remains to be computed.
Using the continuity equation (\ref{genconlaw}), 
$\D_t \rho + \D_x J^x = -\D_y J^y = E.$  
Thus, to find $(\rho, J^x),$ compute $\mathrm{Div}^{-1} E,$ 
where this time the divergence is with respect to $t$ and $x.$
Proceed as in the previous example.
First, substitute (\ref{KPrankneg3coefvalues}) and $c_3 = -2$ into
(\ref{KPdycandidatereplaced}),
\begin{eqnarray}
\label{KPdtycomponent}
E \!\!\! &=& \!\!\! 
t \Big( 2 u + \left( \sigma^2 y^2 - t x \right) 
(u_{tx} + \alpha u_x^2 + \alpha u u_{2x} + u_{4x}) \Big).
\end{eqnarray}
Second, compute the integrands for the homotopy operator, 
\begin{eqnarray}
I_{u(t,x)}^{(t)} E \!\!\! &=& \!\!\!
- \tfrac{1}{2} (u \D_x - u_x \Id) \frac{\pl E}{\pl u_{tx}}
= \tfrac{1}{2} t (t u + (\sigma^2 y^2 - t x ) u_x ), 
\\ 
I_{v(t,x)}^{(t)} E \!\!\! &=& \!\!\! 0,
\\ 
I_{u(t,x)}^{(x)} E \!\!\! &=& \!\!\! u \frac{\pl E}{\pl u_x}
\!-\! \tfrac{1}{2} (u \D_t \!-\! u_t \Id) \frac{\pl E}{\pl u_{tx}}
\!-\! (u \D_x \!-\! u_x \Id) \frac{\pl E}{\pl u_{2x}}
\!-\! (u \D_x^3 \!-\! u_x \D_x^2 \!+\! u_{2x} \D_x
\!-\! u_{3x} \Id) \frac{\pl E}{\pl u_{4x}}
\nonumber \\ 
\!\!\! &=& \!\!\! t^2 (\alpha u^2 + u_{2x})
+ t ( \sigma^2 y^2 - t x ) (\tfrac{1}{2} u_t + 2 \alpha u u_x + u_{3x})
- ( \tfrac{1}{2} \sigma^2 y^2 - t x ) u,
\\ 
I_{v(t,x)}^{(x)} E \!\!\! &=& \!\!\! 0.
\end{eqnarray}
Next, compute
\begin{eqnarray}
\label{rhohatKP}
\hat{\rho} \!\!\! &=& \!\!\!  \mc{H}_{\mf{u}(t,x)}^{(t)} E 
= \int_0^1 \left( I_{u(t,x)}^{(t)} E + I_{v(t,x)}^{(t)} E \right)
[\lambda u] \frac{d \lambda}{\lambda}
\nonumber \\ 
\!\!\! &=& \!\!\! \int_0^1 \left( 
\tfrac{1}{2} t (t u + (\sigma^2 y^2 - t x) u_x ) \right) d\lambda
= \tfrac{1}{2} t (t u + (\sigma^2 y^2 - t x) u_x ), 
\\
\label{JhatKP}
\hat{J}^x \!\!\! &=& \!\!\! \mc{H}_{\mf{u}(t,x)}^{(x)} E 
= \int_0^1 \left( I_{u(t,x)}^{(x)} E + I_{v(t,x)}^{(x)} E \right)
[\lambda u] \frac{d \lambda}{\lambda}
\nonumber \\ 
\!\!\! &=& \!\!\! \int_0^1 
\left( t^2 (\alpha \lambda u^2 + u_{2x})
+ t ( \sigma^2 y^2 - t x ) (\tfrac{1}{2} u_t + 2 \alpha \lambda u u_x + u_{3x})
- (\tfrac{1}{2} \sigma^2 y^2 - t x ) u \right) d\lambda
\nonumber \\ 
\!\!\! 
&=& \!\!\! t^2 ( \tfrac{1}{2} \alpha u^2 + u_{2x})
+ t (\sigma^2 y^2 - t x) (\tfrac{1}{2} u_t + \alpha u u_x + u_{3x})
- (\tfrac{1}{2} \sigma^2 y^2 - t x) u, 
\end{eqnarray}
and remove the curl term $\mf{K} = (\D_x \theta, -\D_t \theta)$ 
with $\theta = \tfrac{1}{2} t ( \sigma ^2 y^2 - t x ) u,$ to obtain
\begin{equation}
\label{KPcomputedconlaw}
\rho =  t^2 u, \quad
J^x =  t^2 (\tfrac{1}{2} \alpha u^2 + u_{2x}) 
+ t (\sigma^2 y^2 - t x)(u_t + \alpha u u_x + u_{3x}).
\end{equation}
The computed conservation law is the same as (\ref{KPgenconlaw1}) where 
$f(t) = t^2$ and $v = u_y.$
%
\section{A Generalized Conservation Law for the KP Equation}
\label{secKPequation}
Due to the presence of an arbitrary function $f(t),$ it is impossible to 
{\it algorithmically} compute (\ref{KPgenconlaw1}) with our code.
The generalization of (\ref{KPcomputedconlaw}) to (\ref{KPgenconlaw1}) is 
based on inspection of the conservation laws in Table~\ref{KPconlawlist} as 
computed by our program {\sc ConservationLawsMD.m}.
%
\begin{table}[h!]
\begin{center}
\begin{tabular}{|r|l|}
\hline Rank & Conservation Law
\\ \hline \hline 5 & $\D_t \Big( u \Big)
+ \D_x \Big( \tfrac{1}{2} \alpha u^2 + u_{2x}
- x (u_t + \alpha uu_x + u_{3x}) \Big)
+ \D_y \Big( \!-\sigma^2 x v \Big) = 0$
\\ \hline 2 & $\D_t \Big( t u \Big)
+ \D_x \Big( t (\tfrac{1}{2} \alpha u^2 + u_{2x})
+ (\tfrac{1}{2} \sigma^2 y^2 - t x) (u_t + \alpha uu_x + u_{3x}) \Big)$
\\ & \hspace*{5mm} ${} + \D_y \Big( (\tfrac{1}{2} y^2
- \sigma^2 t x) v - y u \Big) = 0.$
\\ \hline $-1$ & $\D_t \Big( t^2 u \Big)
+ \D_x \Big( t^2 (\tfrac{1}{2} \alpha u^2
+ u_{2x}) + t (\sigma^2 y^2- t x) (u_t + \alpha uu_x + u_{3x}) \Big)$
\\ & \hspace*{5mm} ${} + \D_y \Big(t (y^2 - \sigma^2 t x) v
-  2 t y u \Big) = 0$
\\ \hline -4 & $\D_t \Big( t^3 u \Big)
+ \D_x \Big( t^3 (\tfrac{1}{2} \alpha u^2 + u_{2x})
+ t^2 (\tfrac{3}{2} \sigma^2 y^2 - t x)
(u_t + \alpha u u_x + u_{3x}) \Big)$
\\ & \hspace*{5mm} ${} + \D_y \Big( t^2 \Big( ( \tfrac{3}{2} y^2
- \sigma^2 t x) v - 3 y u \Big) \Big) = 0$
\\  \hline
\end{tabular}
\end{center}
\caption{
Additional conservation laws for the KP equation (\ref{KPevolutionform}).
}
\label{KPconlawlist}
\end{table}
Indeed, pattern matching with the results in Table~\ref{KPconlawlist} 
and some interactive work lead to (\ref{KPgenconlaw1}), which can be then 
be verified with {\sc ConservationLawsMD.m} as follows.

The conservation laws in Table~\ref{KPconlawlist} suggest that a density has
the form $t^n u,$ or more general, $f(t) u,$ where $f(t)$ is an arbitrary 
function.
The corresponding flux would be harder to guess. 
However, it can be computed as follows.
Since the KP equation (\ref{KPevolutionform}) is an evolution equation in $y,$
we construct a suitable candidate for $J^y.$ 
Guided by the results in Table~\ref{KPconlawlist}, we take 
\begin{equation}
\label{KPformofycomponenet}
J^y = c_1 f^{\prime}(t) y u + c_2 f^{\prime}(t) y^2 v + c_3 f(t) x v,
\end{equation}
where $c_1, c_2,$ and $c_3$ are undetermined coefficients, and $u_y$ is 
replaced by $v$ in agreement with (\ref{KPevolutionform}).
As before, we compute $\D_y J^y$ and replace $u_y$ and $v_y$ using 
(\ref{KPevolutionform}).
Doing so, 
\begin{equation}
\label{KPformofycomponentreplaced}
E = \D_y P^y 
= c_1 f^{\prime} u + (c_1 + 2 c_2)f^{\prime} y v 
- (\sigma^2 c_2 f^{\prime} y^2 + \sigma^2 c_3 f x)
(u_{tx} + \alpha u_x^2 + \alpha u u_{2x} + u_{4x}).
\end{equation}
By (\ref{genconlaw}), $\D_y J^y = - \mathrm{Div} (\rho, J^x).$
By Theorem~\ref{zeroeulerexact},
\begin{equation}
\label{KPvarderformofycomponent}
(0,\, 0) 
= \mf{0} \equiv \mc{L}_{\mf{u}(t,x)} E 
= \left( (c_1 - \sigma^2 c_3) f^{\prime},\, 
         (c_1 + 2 c_2) f^{\prime} y \right).
\end{equation}
Clearly, $c_2 = -\tfrac{1}{2} c_1$ and $c_3 = \sigma^2 c_1.$
If we set $c_1 = -1$ and $v = u_y$ we obtain $J^y$ in (\ref{KPgenconlaw1}).
Application of the homotopy operator (in this case to an expression with 
arbitrary functional coefficients) yields $(\rho, J^x).$ 
This is how conservation law (\ref{KPgenconlaw1}) was computed.
Conservation law (\ref{KPgenconlaw2}) was obtained in a similar way. 
Both conservation laws were then verified using the 
{\sc ConservationLawsMD.m} code.
%
\section{Applications}
\label{secApplications}
In this section we state results obtained by using our algorithm on a variety
of (2+1)-~and (3+1)-dimensional nonlinear PDEs.
The selected PDEs highlight several of the issues that arise when using our
algorithm and software package {\sc ConservationLawsMD.m}.
%
\subsection{The Sawada-Kotera Equation in 2\,D}
\label{secSKequation}
The (2+1)-dimensional SK equation \citep{KonopelchenkoDubrovsky84}, 
\begin{equation}
\label{SKequation}
u_t = 5 u^2 u_x + 5 u u_{3x} + 5 u u_y + 5 u_x u_{2x} + 5 u_{2xy}
+ u_{5x} - 5 \pl_x^{-1}u_{2y} + 5 u_x \pl_x^{-1} u_y,
\end{equation}
with $u(\mf{x}) = u(x,y,t)$ 
is a {\it completely integrable} 2\,D generalization of the standard 
SK equation. 
The latter has infinitely many conservation laws (see, e.g., 
\citet{GoktasHeremanJSC97}).
Our algorithm can not handle the integral terms in (\ref{SKequation}),
so we set $v = \pl_x^{-1} u_y.$  
Doing so, (\ref{SKequation}) becomes a system of evolution equations in $y$:
\begin{equation}
\label{SKsystem}
v_y = - \tfrac{1}{5} u_t + u^2 u_x + u u_{3x} 
      + u v_x + u_x u_{2x} + v_{3x} + \tfrac{1}{5} u_{5x} + u_x v,
\quad u_y = v_x.
\end{equation}
Application of our algorithm to (\ref{SKsystem}) yields several conservation 
laws, all of which have densities $u,$ $t u,$ $t^2 u,$ etc., and 
$y u,$ $t y u,$ $t^2 y u,$ etc.\  
Like with the KP equation, this suggests that there are conservation laws 
with an arbitrary functional coefficient $f(t).$
Proceeding as in Section~\ref{secKPequation} and using 
{\sc ConservationLawsMD.m}, we obtained 
\begin{eqnarray}
\label{2DSKconslaw1}
&&\!\!\!\!\!\!\!\!\!\!\!\!
\D_t \Big(f u \Big) 
\!+\! \D_x \Big( f^{\prime} y v 
\!\!-5 f ( \tfrac{1}{3} u^3 + u v + u u_{2x} + u_{xy} + \tfrac{1}{5} u_{4x}) 
\Big)
\!+\! \D_y \Big( 5 f v - f^{\prime} y u \Big) = 0,
\\  
\label{2DSKconslaw2}
&&\!\!\!\!\!\!\!\!\!\!\!\!
\D_t \Big( f y u \Big) 
\!+\! \D_x \Big( (\tfrac{1}{2} f^{\prime} y^2 - 5 f x) v 
\!\!-5 f y ( \tfrac{1}{3} u^3 + u v + u u_{2x} + u_{xy} +\tfrac{1}{5} u_{4x} ) 
\Big)
\nonumber \\ 
&& \hspace{5mm} {} 
\!+\! \D_y \Big( 5 f y v - (\tfrac{1}{2} f^{\prime} y^2  - 5 f x ) u \Big) = 0.
\end{eqnarray}
Note that the densities in (\ref{2DSKconslaw1}) and (\ref{2DSKconslaw2}) are 
identical to those in (\ref{KPgenconlaw1}) and (\ref{KPgenconlaw2}) for the 
KP equation.
These two densities occur often in (2+1)-dimensional PDEs that have a $u_{tx}$
instead of a $u_t$ term, as shown in the next example.
%
\subsection{The Khokhlov-Zabolotskaya Equation in 2\,D and 3\,D}
\label{secKZequation}
The Khokhlov-Zabolotskaya (KZ) equation or dispersionless KP equation 
describes the propagation of sound in non-linear media in two or three 
space dimensions \citep{SandersWang97}.
The (2+1)-dimensional KZ equation,
\begin{equation}
\label{KZequation2D}
(u_t - u u_x)_x - u_{2y} = 0,
\end{equation}
with $u(\mf{x}) = u(x,y,t)$ can be written as a system of evolution
equations in $y,$ 
\begin{equation}
u_y = v, \quad v_y = u_{tx} - u_x^2 - u u_{2x}, 
\end{equation}
by setting $v = u_y.$ 
Again, two familiar densities appear in the following conservation laws, 
computed indirectly as we showed for the KP and SK equations,
\begin{eqnarray}
\label{2DKZconslaw1}
&& \!\!\!\!\!\!\!\!\!\!\!\! 
\D_t ( u_x ) \!+\! \D_x ( \!-u u_x ) \!+\! \D_y ( \!-u_y ) = 0,
\\ 
\label{2DKZconslaw2}
&& \!\!\!\!\!\!\!\!\!\!\!\! 
\D_t \Big( f u \Big) 
\!+\! \D_x \Big( \!-\!\tfrac{1}{2} f u^2 
\!-\! (\tfrac{1}{2} f^{\prime} y^2 + f x ) (u_t - u u_x) \Big)
\!+\! \D_y \Big( ( \tfrac{1}{2} f^{\prime} y^2 + f x) u_y 
\!-\! f^{\prime} y u \Big) = 0, 
\\ 
\label{2DKZconslaw3}
&& \!\!\!\!\!\!\!\!\!\!\!\! 
\D_t \Big( f y u \Big) 
\!+\! \D_x \Big( \!-\tfrac{1}{2} f y u^2 
\!-\! y ( \tfrac{1}{6} f^{\prime} y^2 + f x) (u_t - u u_x) \Big) 
\nonumber \\ 
&&  
{} \!+\! \D_y \Big( y ( \tfrac{1}{6} f^{\prime} y^2 + f x ) u_y 
\!-\! ( \tfrac{1}{2} f^{\prime} y^2 \!+\! f x ) u \Big) = 0,
\end{eqnarray}
where $f(t)$ is an arbitrary function.
Actually, (\ref{2DKZconslaw2}) and (\ref{2DKZconslaw3}) are {\em nonlocal} 
because, from (\ref{KZequation2D}), 
$u_t - u u_x = \int u_{2y} \, dx.$
By swapping terms in the density and the $x$-component of the flux, 
(\ref{2DKZconslaw2}) with $f(t) = 1,$ can be rewritten as
\begin{equation}
\label{2DKZconlaw2bis}
\D_t \Big(x u_x \Big) + \D_x \Big( \tfrac{1}{2} u^2 - x u u_x \Big)
+ \D_y \Big( \!-x u_y \Big) = 0,
\end{equation}
which is local.
The computation of conservation laws for the (3+1)-dimensional KZ equation,
\begin{equation}
\label{KZequation3D}
(u_t - u u_x)_x - u_{2y} - u_{2z} = 0,
\end{equation}
where $u(\mf{x}) = u(x,y,z,t),$ is more difficult.
This equation can be written as a system of evolution equations in either 
$y$ or $z.$ 
Although the intermediate results differ, either choice leads to equivalent 
conservation laws.
Writing (\ref{KZequation3D}) as an evolution system in $z,$
\begin{equation}
u_z = v, \quad v_z = u_{tx} - u_x^2 - u u_{2x} - u_{2y},
\end{equation}
{\sc ConservationLawsMD.m} is able to compute a variety of conservation laws 
whose densities are shown in Table~\ref{densitiesforKZ3D}.
\vskip 1pt
\noindent
\begin{table}[!h]
\vspace*{-3mm}
\begin{center}
\begin{tabular}{|r|l|} \hline
Rank & Densities Explicitly Dependent on $x,y,z$
\\ \hline \hline 2 & $\rho_1 = x u_x$
\\ \hline \rule[1mm]{0mm}{3mm} 0 & $\rho_2 = x y u_x,$
\hspace*{2mm} $\rho_3 = x z u_x$
\\ \hline \rule[1mm]{0mm}{3mm} $-1$ & $\rho_4 =  t u$
\\ \hline \rule[1mm]{0mm}{3mm} $-2$ & $\rho_5 = x y z u_x,$ \hspace*{2mm}
$\rho_6 = x (y^2-z^2) u_x$
\\ \hline \rule[1mm]{0mm}{3mm} $-3$ & $\rho_7 = t y u,$
\hspace*{2mm} $\rho_8 = t z u$
\\ \hline \rule[1mm]{0mm}{3mm} $-4$ &$\rho_9 = t^2 u,$ \hspace*{2mm}
$\rho_{10} = x y (y^2 - 3 z^2) u_x,$
\hspace*{2mm} $\rho_{11} = x z^2 (3 y - z) u_x$
\\ \hline \rule[1mm]{0mm}{3mm} $-5$ & $\rho_{12} = t y z u,$ \hspace*{2mm}
$\rho_{13} = t (y^2 - z^2) u$
\\ \hline \rule[1mm]{0mm}{3mm} $-6$ & $\rho_{14} = t^2 y u,$ \hspace*{2mm}
$\rho_{15} = t^2 x z u,$
\hspace*{2mm} $\rho_{16} = x y z (y^2 - z^2) u_x,$ 
\hspace*{2mm}
$\rho_{17} = x (y^4 - 6 y^2 z^2 + z^4) u_x$
\\ \hline \rule[1mm]{0mm}{3mm} $-7$ & $\rho_{18} = t^3 u,$ \hspace*{2mm}
$\rho_{19} = t y (y^2 - 3 z^2) u,$
\hspace*{2mm} $\rho_{20} = t z (3 y^2 - z^2) u$
\\ \hline \rule[1mm]{0mm}{3mm} $-8$ & $\rho_{21} = t^2 y z u,$ \hspace*{2mm}
$\rho_{22} = t^2 (y^2 - z^2) u,$
\hspace*{2mm} $\rho_{23} = x y (y^4 - 10 y^2 z^2 + 5 z^4) u_x,$
\\ \rule[1mm]{0mm}{3mm} & $\rho_{24} = x z (5 y^4 - 10 y^2 z^2 + z^4) u_x$
\\ \hline
\end{tabular}
\end{center}
\caption{
Densities for the (3+1)-dimensional KZ equation (\ref{KZequation3D}).
}
\label{densitiesforKZ3D}
\end{table}
\vskip 0.000001pt
\noindent
Density $\rho_1 = x u_x$ in Table~\ref{densitiesforKZ3D} is part of 
local conservation law
\begin{equation}
\label{3DKZconlawrank2}
\D_t \Big(x u_x \Big) + \D_x \Big( \tfrac{1}{2} u^2 - x u u_x \Big)
+ \D_y \Big( \!-x u_y \Big) + \D_z \Big(\!- x u_z \Big) = 0,
\end{equation}
which can be rewritten as a nonlocal conservation law
\begin{equation}
\label{3DKZconlawrank2rev}
\D_t \Big( u \Big) + \D_x \Big( \!-\tfrac{1}{2} u^2 - x (u_t - u u_x) \Big)
+ \D_y \Big( x u_y \Big) + \D_z \Big(x u_z \Big) = 0.
\end{equation}
In general, if a factor $x u_x$ appears in a density then that factor can 
be replaced by $u.$ 
Doing so, all densities in Table~\ref{densitiesforKZ3D} that can be 
expressed as $\rho = g(y,z,t) u,$ where $g(y,z,t)$ is arbitrary.
Introducing an arbitrary function $h = h(y,z,t),$ the conservation laws 
corresponding to the densities in Table~\ref{densitiesforKZ3D} can be 
summarized as
\begin{eqnarray}
\label{KZgenconlaw3} \!\!\!\!\!\!\!\!\!\!
&& \D_t \Big( g u \Big) \!+\! \D_x \Big( \!-\tfrac{1}{2} g u^2
- (x g + h) (u_t - u u_x) \Big) \!+\! \D_y \Big( ( x g + h) u_y
- (x g_y + h_y) u \Big) \nonumber
\\ \!\!\!\!\!\!\!\!\!\! 
&& \hspace*{5mm} {}
\!\!\!\!\!\!\!\! \!+\! \D_z \Big( (x g + h) u_z - (x g_z + h_z) u \Big)
=  - \Big( h_{2y} + h_{2z} - g_t + x ( g_{2y} + g_{2z} ) \Big) u.
\end{eqnarray}
Equation (\ref{KZgenconlaw3}) is only a conservation law when the 
constraints $\Delta g = 0$ and $\Delta h = g_t$ are satisfied, 
where $\Delta = \frac{\pl^2}{\pl y^2} + \frac{\pl^2}{\pl z^2}.$
Thus, $g$ must be a harmonic function and $h$ must satisfy the Poisson 
equation with $g_t$ on the right hand side.
Combining both equations produces the biharmonic equation 
$\Delta^2 h = 0.$ 
As shown by \citet{TikhonovSamarskii63}, $\Delta^2 h = 0$ has general 
solutions of the form
\begin{equation}
\label{biharmnonicsol1and2}
h = y \, h_1(y,z) + h_2(y,z) \quad {\rm and} \quad
h = z \, h_1(y,z) + h_2(y,z),
\end{equation}
where $\Delta h_1 = 0$ and $\Delta h_2 = 0.$
Treating $t$ as a parameter, four solutions for $h(y,z,t)$ are
\begin{eqnarray}
\label{relategtof1}
h(y,z,t) \!\!\! &=& \!\!\! \tfrac{1}{2} y \, \pl_y^{-1} g_t(y,z,t),
\\ 
\label{relategtof2}
h(y,z,t) \!\!\! &=& \!\!\! \tfrac{1}{2} \pl_y^{-1}(y g_t)
= \tfrac{1}{2} (y \, \pl_y^{-1} g_t(y,z,t) - \pl_y^{-2} g_t(y,z,t)),
\\ 
\label{relategtof3}
h(y,z,t) \!\!\! &=& \!\!\! \tfrac{1}{2} z \, \pl_z^{-1} g_t(y,z,t),
\\ 
\label{relategtof4}
h(y,z,t) \!\!\! &=& \!\!\! \tfrac{1}{2} \pl_z^{-1}(z g_t)
= \tfrac{1}{2} (z \, \pl_z^{-1} g_t(y,z,t) - \pl_z^{-2} g_t(y,z,t)).
\end{eqnarray}
This shows how $h$ can be written in terms of $g.$
For every conservation law corresponding to the densities in 
Table~\ref{densitiesforKZ3D}, $h$ could be computed using one of the 
equations in (\ref{relategtof1})-(\ref{relategtof4}).

Conservation laws for the KZ equation have been reported in the literature 
by \citet{Sharomet89} and \citet{SandersWang97}.
However, substitution of their results into (\ref{divconlaw}) revealed
inaccuracies.
After bringing the mistake to their attention, 
\citet{SandersWang97Revised} have since corrected one of their conservation 
laws to match our result.
%
\subsection{The Camassa-Holm Equation in 2\,D}
\label{secCHequation}
The (2+1)-dimensional CH equation,
\begin{equation}
\label{CHequation}
(u_t + \kappa u_x - u_{t2x} + 3 u u_x - 2 u_x u_{2x} - u u_{3x})_x
+ u_{2y} = 0,
\end{equation}
for $u(\mf{x}) = u(x,y,t)$ models water waves \citep{Johnson02}.
It is an extension of the completely integrable 1\,D CH equation 
derived by \citet{CamassaHolm93}.
A study by \citet{GordoaPickeringSenthilvelan04} concluded that 
(\ref{CHequation}) is not completely integrable.

Obviously, (\ref{CHequation}) is a conservation law itself,
\begin{equation}
\label{CHequationConslaw}
\D_t ( u_x - u_{3x} ) 
+ \D_x ( \kappa u_x + 3 u u_x - 2 u_x u_{2x} - u u_{3x} )
+ \D_y ( u_y ) = 0. 
\end{equation}
It can be written as a system of evolution equations in $y.$ 
Indeed,
\begin{equation}
\label{CHevolutiony}
u_y = v, \quad
v_y =  - (\alpha u_{t} + \kappa u_{x} - u_{t2x} + 3 \beta u u_{x} 
- 2 u_x u_{2x} - u u_{3x})_x.
\end{equation}
Note that we introduced auxiliary parameters $\alpha$ and $\beta$ as 
coefficients of the $u_t$ and $u u_x$ terms, respectively.
The reason for doing so is that the CH equation (\ref{CHequation}) does not 
have a scaling symmetry unless we add scales on the parameters 
$\alpha, \beta$ and $\kappa.$
Our code guided us in finding the following conservation laws with 
functional coefficients,
\begin{eqnarray}
\!\!\!\!\!\!\!\!\!\!\!\!\!
&& \D_t \Big( f u \Big)
\!+\! \D_x \Big( \tfrac{1}{\alpha} f (\tfrac{3}{2} \beta u^2 
\! + \kappa u - \tfrac{1}{2} u_x^2 \! - u u_{2x} - u_{tx})
+ (\tfrac{1}{2} f^{\prime} y^2 \! - \tfrac{1}{\alpha} f x)
(\alpha u_t \! + \kappa u_x
\nonumber \\ 
\!\!\!\!\!\!\!\!\!\!\!\!\! 
&& \hspace*{1mm} 
{} + 3 \beta u u_x - 2 u_x u_{2x} - u u_{3x} - u_{t2x}) \Big)
\!+\! \D_y \Big( (\tfrac{1}{2} f^{\prime} y^2 - \tfrac{1}{\alpha} f x ) u_y 
- f^{\prime} y u \Big) = 0,
\\ 
\!\!\!\!\!\!\!\!\!\!\!\!\!
&& \D_t \Big( f y u \Big) \!+\! \D_x \Big( \tfrac{1}{\alpha} f y
(\tfrac{3}{2} \beta u^2 \!+\! \kappa u \!-\! \tfrac{1}{2} u_x^2 
\!-\! u u_{2x} \!-\! u_{tx})
\!+\! y (\tfrac{1}{6} f^{\prime} y^2 \!-\! \tfrac{1}{\alpha} f x  ) 
(\alpha u_t \!+\! \kappa u_x
\nonumber \\ 
\!\!\!\!\!\!\!\!\!\!\!\!\! 
&& \hspace*{1mm} 
{} \!+\! 3 \beta u u_x \! \!-\! 2 u_x u_{2x} \!-\! u u_{3x} \!-\! u_{t2x})\Big)
\!+\! \D_y \Big( 
y (\tfrac{1}{6} f^{\prime} y^2 \!-\! \tfrac{1}{\alpha} f x ) u_y
+ (\tfrac{1}{\alpha} f x \!-\! \tfrac{1}{2} f^{\prime} y^2 ) u
\Big) = 0,
\end{eqnarray}
where $f(t)$ is arbitrary and without constraints on the parameters.
Thus, if we set $\alpha = \beta = 1,$ we have conservation laws for 
(\ref{CHequation}).

\subsection{The Gardner Equation in 2\,D}
\label{secGardnerequation}
The (2+1)-dimensional Gardner equation \cite{KonopelchenkoDubrovsky84}, 
\begin{equation}
\label{Gardnerequationorig}
u_t = -\tfrac{3}{2} \alpha^2 u^2 u_x + 6 \beta u u_x + u_{3x} 
- 3 \alpha u_x \pl_x^{-1}u_y + 3 \pl_x^{-1} u_{2y},
\end{equation}
for $u(\mf{x}) = u(x,y,t)$ is a 2\,D generalization of 
\begin{equation}
\label{GardnerEqn1D}
u_t = - \tfrac{3}{2} \alpha u^2 u_x + 6 \beta u u_x + u_{3x},
\end{equation}
which is an integrable combination of the KdV and mKdV equations due 
to Gardner.
For $\alpha = 0,$ (\ref{Gardnerequationorig}) reduces to the KP equation
(\ref{KPequation}).
For $\beta = 0,$ (\ref{Gardnerequationorig}) becomes a modified KP equation. 
Adding a new dependent variable, $v = \pl_x^{-1} u_y,$ allows one to remove 
the integral terms and replace (\ref{Gardnerequationorig}) by the system
\begin{equation}
\label{Gardnersystemorig}
u_y = v_x, \quad
v_y = \tfrac{1}{3} u_t -\tfrac{1}{3} u_{3x} - 2 \beta u u_x + \alpha u_x v 
+ \tfrac{1}{2}\alpha^2 u^2 u_x.
\end{equation}
For (\ref{Gardnerequationorig}), we found two conservation laws with 
constant coefficients, 
\begin{eqnarray} 
\!\!\!\!\!\!\!\!\!\!\!\!\!\!\!\!\!\!\!\!
\label{GardnerCLpoly1}
&& \D_t \Big(u \Big) + \D_x \Big( \tfrac{1}{2} \alpha^2 u^3 - 3 \beta u^2
+ 3 \alpha u v - u_{2x} \Big) 
+ \D_y \Big( - ( \tfrac{3}{2} \alpha u^2 + 3 v ) \Big) = 0, 
\\
\label{GardnerCLpoly2} 
\!\!\!\!\!\!\!\!\!\!\!\!\!\!\!\!\!\!\!\!
&& \D_t \Big( u^2 \Big) + \D_x \Big( \tfrac{3}{4} \alpha^2 u^4
- 4 \beta u^3 + 3\alpha u^2 v + 3v^2 + u_x^2 - 2 u u_{2x} \Big)
+ \D_y \Big( - u (\alpha u^2 + 6 v) \Big) = 0.
\end{eqnarray}
Using the methodology described for the previous examples in this section, 
we eventually found three conservation laws involving a variable 
coefficient $f(t),$
\begin{eqnarray}
\label{Gardnergenconlaw1} 
\!\!\!\!\!\!\!\!\!\!
&& \D_t \Big( f u \Big) + \D_x \Big( f ( \tfrac{1}{2} \alpha^2 u^3
- 3 \beta u^2 + 3 \alpha u v - u_{2x}) + f^{\prime} y v \Big)
\nonumber
\\ 
\!\!\!\!\!\!\!\!\!\! 
&& \hspace{5mm} {}
+ \D_y \Big( - f (\tfrac{3}{2} \alpha u^2 + 3 v) - f^{\prime} y u \Big) = 0,
\\
\label{Gardnergenconlaw2} 
\!\!\!\!\!\!\!\!\!\!
&& \D_t \Big( u ( f u + \tfrac{2}{3 \alpha} y f^{\prime} ) \Big)
+ \D_x \Big( f (\tfrac{3}{4} \alpha^2 u^4 - 4 \beta u^3 + 3 \alpha u^2 v
+ 3 v^2 + u_x^2 - 2 u u_{2x}) 
\nonumber \\ 
\!\!\!\!\!\!\!\!\!\! 
&& \hspace{5mm} {} 
+ \tfrac{2}{3 \alpha} y f^{\prime} (\tfrac{1}{2} \alpha^2 u^3 - 3 \beta u^2 
+ 3 \alpha u v - u_{2x}) 
+ \tfrac{1}{\alpha} (2 x f^{\prime} 
+ \tfrac{1}{3} y^2 f^{\prime\prime} ) v \Big) 
\nonumber \\ 
\!\!\!\!\!\!\!\!\!\! 
&& \hspace{5mm} {} 
+ \D_y \Big( 
- f u ( \alpha u^2 + 6 v) 
- \tfrac{1}{\alpha} y f^{\prime} ( \alpha u^2 + 2 v )
- \tfrac{1}{\alpha} ( \tfrac{1}{3} y^2 f^{\prime\prime} + 2 x f^{\prime} ) u  
\Big) = 0,
\end{eqnarray}
and
\begin{eqnarray}
\label{Gardnergenconlaw3} 
\!\!\!\!\!\!\!\!\!\!
&& \D_t \Big( ( \tfrac{\alpha}{6} y f^{\prime} + \beta f ) u^2
+ \tfrac{1}{3} ( \tfrac{1}{6} y^2 f^{\prime\prime} + x f^{\prime} ) u \Big)
+ \D_x \Big( ( \tfrac{\alpha}{6} y f^{\prime} + \beta f )
( \tfrac{3}{4} \alpha^2 u^4 - 4 \beta u^3 
\nonumber \\ 
\!\!\!\!\!\!\!\!\!\! 
&& \hspace{5mm} {} 
+ 3 \alpha u^2 v + 3 v^2 + u_x^2 - 2 u u_{2x} ) 
+ \tfrac{1}{3} ( \tfrac{1}{6} y^2 f^{\prime\prime} + x f^{\prime} ) 
(\tfrac{1}{2} \alpha^2 u^3 - 3 \beta u^2 + 3 \alpha u v - u_{2x})
\nonumber \\ 
\!\!\!\!\!\!\!\!\!\! 
&& \hspace{5mm} {} 
+ \tfrac{1}{3} f^{\prime} u_x 
+ \tfrac{1}{3} y 
( \tfrac{1}{18} y^2 f^{\prime\prime\prime} + x f^{\prime\prime} ) v \Big)
+ \D_y \Big( -(\tfrac{\alpha}{6} y f^{\prime} + \beta f )(\alpha u^2 + 6 v ) u
\nonumber \\ 
\!\!\!\!\!\!\!\!\!\! 
&& \hspace{5mm} {} 
- \tfrac{1}{2} ( \tfrac{1}{6} y^2 f^{\prime\prime} + x f^{\prime} ) 
  (\alpha u^2 + 2 v) 
- \tfrac{1}{3} y 
( \tfrac{1}{18} y^2 f^{\prime\prime\prime} + x f^{\prime\prime} ) u \Big) = 0.
\end{eqnarray}
Setting $f(t) = 1$ in (\ref{Gardnergenconlaw1}) and (\ref{Gardnergenconlaw2})
yields (\ref{GardnerCLpoly1}) and (\ref{GardnerCLpoly2}), respectively.
%
\section{Using the Program ConservationLawsMD.m}
\label{secUsingProgram}
Before using {\sc ConservationLawsMD.m}, all data files provided with the 
program, as well as additional data files created by the user, must be placed 
into one directory.
Next, open the {\it Mathematica} notebook {\sc ConservationLawsMD.nb} which 
contains instructions for loading the code.
Executing the command {\sc ConservationLawsMD[]} will open a menu, offering 
the choice of computing conservation laws for a PDE from the menu or from 
a data file prepared by the user.
All PDEs listed in the menu have matching data files.
An example of a data file is shown in Figure~\ref{dataforGardner}.

The independent space variables must be $x,$ $y,$ and $z.$ 
The symbol $t$ must be used for time. 
Dependent variables must be entered as $u_i,$ $i = 1, \dots, N,$
where $N$ is the number of dependent variables.
In a (1+1)-dimensional case, the dependent variables 
(in {\it Mathematica} syntax) are {\tt u[1][x,t]}, {\tt u[2][x,t]}, etc.\
In a (3+1)-dimensional cases, {\tt u[1][x,y,z,t], u[2][x,y,z,t]}, etc.,
where $t$ is always the last argument. 
%
\section{Conclusions}
\label{secConclusion}
We have presented an algorithm and a software package, 
{\sc ConservationLawsMD.m}, to compute conservation laws of nonlinear 
polynomial PDEs in multiple space dimensions.

In contrast to the approach taken by researchers working with {\it Maple} 
and {\it Reduce}, our algorithm uses only tools from calculus, the calculus 
of variations, linear algebra, and differential geometry.
In particular, we do {\it not} first compute the determining PDEs for the 
density and the flux components and then attempt to solve these PDEs.
Although restricted to polynomial conservation laws, our constructive method 
leads to short densities (free of divergences and divergence-equivalent terms)
and curl-free fluxes.

The software is easy to use, runs fast, and has been tested for a variety of 
multi-dimensional nonlinear PDEs, demonstrating the versatility of the code.
Many of the test cases have been added to the menu of the program.
In addition, the program allows the user to test conservation laws either 
computed with other methods, obtained from the literature, or conjectured 
after work with the code.
The latter is particularly relevant for finding conservation laws involving 
arbitrary functions as shown in Sections~\ref{secKPequation} 
and~\ref{secApplications}.

Currently, {\sc ConservationLawsMD.m} has two major limitations:
(i) the PDE must either be an evolution equation or correspond to a system 
of evolution equations, perhaps after an interchange of independent 
variables or some other transformation; and 
(ii) the program can only generate local polynomial densities and fluxes. 
However, the {\it testing} capabilities of {\sc ConservationLawsMD.m} 
are more versatile.
The code can be used to test conservation laws involving smooth functions of 
the independent variables and the densities and fluxes are not restricted to 
polynomial differential functions.  

Future versions of the code will work with any number of independent 
variables and will cover PDEs that are not of evolution type, e.g., 
PDEs with mixed derivatives and transcendental nonlinearities.
\vskip 1pt
\noindent
\begin{figure}[hbt]
{\tt (* data file d\_kd2d.m *)}
\\ {\tt (* Menu item 2-10 *)}
\vskip 6pt
{\tt (*** 2\,D Gardner equation from \citet{KonopelchenkoDubrovsky84} ***)}
\vskip 6pt
{\tt eq[1] \!=\! D[u[1][x,y,t],y] \!-\! D[u[2][x,y,t],x];}
\vskip 6pt
{\tt eq[2] \!=\! D[u[2][x,y,t],y] -(1/3)*D[u[1][x,y,t],t] 
+ (1/3)*D[u[1][x,y,t],{x,3}] 
\\ 
\hspace*{1mm}  
\!+\!2*beta*u[1][x,y,t]*D[u[1][x,y,t],x] 
\!-\!alpha*D[u[1][x,y,t],x]*u[2][x,y,t]
\\ 
\hspace*{1mm} 
-(1/2)*alpha$^{\wedge}$2*u[1][x,y,t]$^{\wedge}$2*D[u[1][x,y,t],x];
}
\vskip 6pt
{\tt diffFunctionListINPUT = \{eq[1],eq[2]\};}
\\ {\tt numDependentVariablesINPUT = 2;} 
\\ {\tt independentVariableListINPUT = \{x,y\};} 
\\ \hspace*{1cm} \parbox{14.5cm}{\rm The space variables only; ignore t.}
\\ {\tt nameINPUT = "(2+1)-dimensional Gardner equation";}
\\ {\tt noteINPUT = "{\rm Any additional information can be put here.}";}
\vskip 6pt
{\tt parametersINPUT = \{alpha\};}
\\ \hspace*{1cm} {\rm All parameters without scaling must be placed in this 
list.}
\\ {\tt weightedParametersINPUT = \{beta\};}
\\ \hspace*{1cm} \parbox{14.5cm}{\rm Parameters that should have a scaling 
factor must be placed in this list.}
\vskip 6pt
{\tt userWeightRulesINPUT = \{\};}
\\ \hspace*{1cm} {\rm Optional: the user can choose scales for variables.}
\\ {\tt rankRhoINPUT = Null;}
\vskip 1pt
\hspace*{1cm} \parbox{14.5cm}{\rm Can be changed to a list of values if the 
user wishes to work with several ranks at once.
The program runs automatically when such values are given.}
\vskip 1pt
{\tt explicitIndependentVariablesInDensitiesINPUT = Null;}
\vskip 1pt
\hspace*{1cm} \parbox{14.5cm}{\rm Can be set to $0, 1, 2, \dots,$  
specifying the maximum degree $(m+n+p)$ of coefficients 
$c_i \, t^m x^n y^p $ in the density.}
\vskip 1pt
{\tt formRhoINPUT = \{\};}
\vskip 1pt
\hspace*{1cm} \parbox{14.5cm}{\rm The user can give a density to be tested. 
However, this works only for evolution equations in variable $t.$}
\vskip 6pt
{\tt (* end of data file d\_kd2d.m *)}
\caption{
Data file for the 2\,D Gardner equation in (\ref{Gardnerequationorig}).
}
\label{dataforGardner}
\vspace*{-2mm}
\end{figure}
%
\section*{Acknowledgements}
\label{secAcknowledgements}
This material is based in part upon work supported by the National Science
Foundation (NSF) under Grant No.\ CCF-0830783.
Any opinions, findings, and conclusions or recommendations expressed in this
material are those of the authors and do not necessarily reflect the views 
of NSF.

Mark Hickman (University of Canterbury, Christchurch, New Zealand) and 
Bernard Deconinck (University of Washington, Seattle) are gratefully
acknowledged for valuable discussions.
Undergraduate students Jacob Rezac, John-Bosco Tran, and Travis ``Alan" Volz 
are thanked for their help with this project.
We thank the anonymous referees whose constructive comments and suggestions 
helped us further improve the manuscript.
\vspace{-2.5mm}
\noindent
%

%

\begin{thebibliography}{99}


\bibitem[{Ablowitz and Clarkson(1991)}]{AblowitzClarkson91}
Ablowitz, M.J., Clarkson, P.A., 1991.
Solitons, Nonlinear Evolution Equations and Inverse Scattering,
Cambridge University Press, Cambridge, U.K.

\bibitem[{Ablowitz and Segur(1981)}]{AblowitzSegur81}
Ablowitz, M.A., Segur, H., 1981.
Solitons and the Inverse Scattering Transform,
SIAM Stud.\ in Appl.\ Math., vol.\ 4, SIAM, Philadelphia, Pennsylvania.

\bibitem[{Anderson(2004a)}]{Anderson04}
Anderson, I.M., 2004a.
The Variational Complex.
Dept.\ of Mathematics, Utah State University, Logan, Utah, 
318 pages, manuscript available at
http://www.math.usu.edu/$\sim$fg\_mp/Publications/VB/vb.pdf.

\bibitem[{Anderson(2004b)}]{Andersonsoftware04}
Anderson, I.M., 2004b.
The Vessiot package; the software with documentation is available at
\newline
\noindent
http://www.math.usu.edu/$\sim$fg\_mp/Pages/SymbolicsPage/VessiotDownloads.html.

\bibitem[{Anderson and Cheb-Terrab(2009)}]{AndersonChebTerMaple09}
Anderson, I.M., Cheb-Terrab, E., 2009.
DifferentialGeometry package, Maple Online Help,
\newline
\noindent
www.maplesoft.com/support/help/Maple/view.aspx?path=DifferentialGeometry.


\bibitem[{Baldwin and Hereman(2010)}]{BaldwinHereman10}
Baldwin, D., Hereman, W., 2010.
A symbolic algorithm for computing recursion operators of nonlinear PDEs,
Int.\ J.\ Comp.\ Math.\ 87, 1094--1119.

\bibitem[{Bluman {\it et al.}(2010)}]{Blumanetal10}
Bluman, G.W., Cheviakov, A.F., Anco, S.C., 2010.
Applications of Symmetry Methods to Partial Differential Equations,
Appl.\ Math.\ Sciences, vol.\ 168, Springer Verlag, New York.


\bibitem[{Camassa and Holm(1993)}]{CamassaHolm93}
Camassa, R., Holm, D.D., 1993.
An integrable shallow water equation with peaked solutions,
Phys.\ Rev.\ Lett.\ 71, 1661--1664.

\bibitem[{Cheb-Terrab and von Bulow(2004)}]{ChebTerrabvonBulowMaple04}
Cheb-Terrab, E., and von Bulow, K., 2004.
PDEtools package, Maple Online Help,
\newline
\noindent
http://www.maplesoft.com/support/help/Maple/view.aspx?path=PDEtools.

\bibitem[{Cheviakov(2007)}]{Cheviakov07}
Cheviakov, A.F., 2007.
GeM software package for computation of symmetries and conservation
laws of differential equations,
Comp.\ Phys.\ Commun.\ 76, 48--61.

\bibitem[{Cheviakov(2010)}]{Cheviakov10}
Cheviakov, A.F., 2010.
{\rm Computation of fluxes of conservation laws}, 
J.\ Engr.\ Math.\ 66, 153--173.


\bibitem[{Deconinck and Nivala(2009)}]{DeconinckNivala09}
Deconinck, B., Nivala, M., 2009.
{\rm Symbolic integration and summation using homotopy operators},
Math.\ Comput.\ Simul.\ 80, 825--836.

\bibitem[{Drinfel'd and Sokolov(1985)}]{DrinfeldSokolov85}
Drinfel'd, V.G., Sokolov, V.V., 1985.
{\rm Lie algebras and equations of Korteweg-de Vries type}, 
J.\ Sov. Math.\ 30, 1975--2036.


\bibitem[{G\"{o}kta\c{s} and Hereman(1997)}]{GoktasHeremanJSC97}
G\"{o}kta\c{s}, \"{U}., Hereman, W., 1997,
Symbolic computation of conserved densities for systems of nonlinear
evolution equations,
J.\ Symbolic Comput.\ 24, 591--621.

\bibitem[{Gordoa {\it et al.}(2004)}]{GordoaPickeringSenthilvelan04}
Gordoa, P.G., Pickering, A., Senthilvelan, M., 2004.
Evidence for the nonintegrability of a water wave equation in 2+1
dimensions,
Zeit.\ f\"{u}r Naturfor.\ 59a, 640--644.


\bibitem[{Hereman(2006)}]{Hereman06}
Hereman, W., 2006.
{\rm Symbolic computation of conservation laws of nonlinear partial 
differential equations in multi-dimensions},
Int.\ J.\ Quant.\ Chem.\ 106, 278--299.

\bibitem[{Hereman {\it et al.}(2008)}]{HeremanAdamsAll08}
Hereman, W., Adams, P.J., Eklund, H.L.,
Hickman, M.S., Herbst, B.M., 2009.
Direct methods and symbolic software for conservation laws of nonlinear
equations. In: Yan, Z. (Ed.), Advances in Nonlinear Waves and Symbolic
Computation, Nova Science Publishers, New York, pp.\ 19--79.

\bibitem[{Hereman {\it et al.}(2005)}]{HeremanAll05}
Hereman, W., Colagrosso, M., Sayers, R., Ringler, A., Deconinck, B.,
Nivala, M., Hickman, M.S., 2005.
Continuous and discrete homotopy operators and the computation
of conservation laws.
In: Wang, D., Zheng, Z. (Eds.), 
Differential Equations with Symbolic Computation, 
Birkh\"{a}user, Basel, pp.\ 249--285.

\bibitem[{Hereman {\it et al.}(2007)}]{Heremanetal07}
Hereman, W., Deconinck, B., Poole, L.D., 2007.
{\rm Continuous and discrete homotopy operators: A theoretical approach made 
concrete},
Math.\ Comput.\ Simul.\ 74, 352--360.


\bibitem[{Infeld(1985)}]{Infeld85}
Infeld, E., 1985.
Self-focusing nonlinear waves,
J.\ Plasma Phys.\ 33, 171--182.


\bibitem[{Johnson(2002)}]{Johnson02}
Johnson, R.S., 2002.
Camassa-Holm, Korteweg-de Vries and related models for water waves,
J.\ Fluid Mech.\ 455, 63--82.


\bibitem[{Kadomtsev and Petviashvili(1970)}]{Kadomtsev70}
Kadomtsev, B.B., Petviashvili, V.I., 1970.
On the stability of solitary waves in weakly dispersive media,
Sov.\ Phys.\ Dokl.\ 15, 539--541.

\bibitem[{Konopelchenko and Dubrovsky(1984)}]{KonopelchenkoDubrovsky84}
Konopelchenko, B.G., Dubrovsky, V.G., 1984.
Some new integrable nonlinear evolution equations in 2+1 dimensions,
Phys.\ Lett.\ A 102, 15--17.


\bibitem[{Lax(1968)}]{Lax68}
Lax, P.D., 1968.
Integrals of nonlinear equations of evolution and solitary waves,
Commun.\ Pure Appl.\ Math.\ 21, 467--490.


\bibitem[Miura {\it et al.}(1968)]{MiuraGardnerKruskal68}
Miura, R.M., Gardner, C.S., Kruskal, M.D., 1968.
Korteweg-de Vries equation and generalizations II. Existence of
conservation laws and constants of motion,
J.\ Math.\ Phys., 9, 1204--1209.


\bibitem[{Naz(2008)}]{Naz08}
Naz, R., 2008.
Symmetry solutions and conservation laws for some partial differential
equations in field mechanics,
Ph.D.\ dissertation, University of the Witwatersrand, Johannesburg.

\bibitem[{Naz {\it et al.}(2008)}]{Nazetal08}
Naz, R., Mahomed, F.M., Mason, D.P., 2008.
Comparison of different approaches to conservation laws for some partial
differential equations in fluid mechanics,
Appl.\ Math.\ Comput.\ 205, 212--230.

\bibitem[{Newell(1983)}]{Newell83}
Newell, A.C., 1983.
The history of the soliton,
J.\ Appl.\ Mech.\ 50, 1127--1138.


\bibitem[{Olver(1993)}]{Olver93}
Olver, P.J., 1993.
Applications of Lie Groups to Differential Equations, 2nd.\ ed.,
Grad.\ Texts in Math., vol.\ 107, Springer Verlag, New York.


\bibitem[{Poole(2009)}]{Poole09}
Poole, L.D., 2009.
Symbolic computation of conservation laws of nonlinear partial differential
equations using homotopy operators,
Ph.D.\ dissertation, Colorado School of Mines, Golden, Colorado.

\bibitem[Poole and Hereman(2009)]{PooleHereman09software1}
Poole, D., Hereman, W., 2009.
{\sc HomotopyIntegrator.m}:
A {\it Mathematica} package for the application of the homotopy method for
(i) integration by parts of expressions involving unspecified functions of
one variable and (ii) the inversion of a total divergence involving
unspecified functions of two or three independent variables;
software available at
http://inside.mines.edu/$\sim$whereman under scientific software.

\bibitem[Poole and Hereman(2009)]{PooleHereman09software2}
Poole, D., Hereman, W., 2009.
{\sc ConservationLawsMD.m}:
A {\it Mathematica} package for the symbolic computation of conservation laws 
of polynomial systems of nonlinear PDEs in multiple space dimensions,
software available at
http://inside.mines.edu/$\sim$whereman under scientific software.

\bibitem[{Poole and Hereman(2010)}]{PooleHereman10}
Poole, D., Hereman, W., 2010.
The homotopy operator method for symbolic integration by parts and inversion
of divergences with applications,
Appl.\ Anal.\ 87, 433--455.


\bibitem[{Rosenhaus(2002)}]{Rosenhaus02}
Rosenhaus V., 2002.
Infinite symmetries and conservation laws, 
J.\ Math.\ Phys.\ 43, 6129--6150.


\bibitem[{Sanders and Wang(1997a)}]{SandersWang97}
Sanders, J., Wang, J.P., 1997a.
Hodge decomposition and conservation laws,
Math.\ Comput.\ Simul.\ 44, 483--493.

\bibitem[{Sanders and Wang(1997b)}]{SandersWang97Revised}
Sanders, J., Wang, J.P., 1997b.
Hodge decomposition and conservation laws; corrected paper, see URL
http://www.math.vu.nl/~jansa/\#research.

\bibitem[{Sanz-Serna(1982)}]{Sanzserna82}
Sanz-Serna, J.M., 1982.
An explicit finite-difference scheme with exact conservation properties,
J.\ Comput.\ Phys.\ 47, 199--210.

\bibitem[{Sharomet(1989)}]{Sharomet89}
Sharomet, N.O., 1989.
Symmetries, invariant solutions and conservation laws of the nonlinear
acoustics equation,
Acta Appl.\ Math.\ 15, 83--120.

\bibitem[{Shivamoggi {\it et al.}(1993)}]{Shivamoggi93}
Shivamoggi, B.K., Rollins, D.K., Fanjul, R., 1993.
Analytic aspects of the Zakharov-Kuznetsov equation,
Phys.\ Scripta 47, 15--17.


\bibitem[{Tikhonov and Samarskii(1963)}]{TikhonovSamarskii63}
Tikhonov, A.N., Samarskii, A.A., 1963.
Equations of Mathematical Physics, Dover Publications, New York.


\bibitem[{Vinogradov(1989)}]{Vinogradov89}
Vinogradov, A.M., 1989.
Symmetries and Conservation Laws of Partial Differential Equations:
Basic Notions and Results,
Acta Appl.\ Math.\ 15, 3--21.


\bibitem[{Wolf(2002)}]{Wolf02}
Wolf, T., 2002.
A comparison of four approaches to the calculation of conservation laws,
Europ.\ J.\ Appl.\ Math.\ 13, 129--152.


\bibitem[{Zakharov and Kuznetsov(1974)}]{ZakharovKuznetsov74}
Zakharov, V.E., Kuznetsov, E.A., 1974.
Three-dimensional solitons,
Sov.\ Phys.\ JETP 39, 285--286.

\bibitem[{Zakharov and Shabat(1972)}]{ZakharovShabat72}
Zakharov, V.E., Shabat, A.B., 1972.
Exact theory of two-dimensional self-focusing and one-dimensional 
self-modulation of waves in nonlinear media, 
Sov.\ Phys.\ JETP 34, 62--69.

\end{thebibliography}
\end{document}